%% file: main.tex
\documentclass[11pt]{article}

\usepackage[utf8]{inputenc}
\usepackage[T1]{fontenc}
\usepackage[margin=1in]{geometry}
\usepackage{lmodern}
\usepackage{microtype}
\usepackage{amsmath,amssymb}
\usepackage[table]{xcolor}
\usepackage{graphicx}
\usepackage{float}
\usepackage{capt-of}
\usepackage{pgfplots}
\pgfplotsset{compat=1.18}
\usepackage{booktabs}
\usepackage{multirow}
\usepackage{makecell}
\usepackage{tabularx}
\usepackage{listings}
\usepackage{caption}
\usepackage{subcaption}
\usepackage{enumitem}
\usepackage[section]{placeins}

\usepackage{tikz}
\usetikzlibrary{arrows.meta,positioning,fit,backgrounds,calc,shapes.misc,shapes.geometric,decorations.pathmorphing}
\usepackage[numbers,sort&compress]{natbib}
\usepackage{xurl}
\usepackage[colorlinks=true,linkcolor=nvgreendark,citecolor=nvgreendark,urlcolor=nvgreendark]{hyperref}
\usepackage[capitalise]{cleveref}

\graphicspath{{figures/}}

\definecolor{nvgreen}{HTML}{76B900}
\definecolor{nvgreendark}{HTML}{4E7A00}
\definecolor{nvgreenlight}{HTML}{E8F5CC}
\definecolor{nvblack}{HTML}{1A1A1A}
\definecolor{nvink}{HTML}{4A4A4A}
\definecolor{nvmuted}{HTML}{8A8A8A}
\definecolor{nvpanel}{HTML}{F4F4F2}
\definecolor{nvgrid}{HTML}{E6E6E6}
\definecolor{nvblue}{HTML}{0067C5}
\definecolor{nvbluelight}{HTML}{E0EDF8}
\definecolor{nvorange}{HTML}{EF9100}
\definecolor{nvorangelight}{HTML}{FDF0DC}
\definecolor{nvpurple}{HTML}{9B3FD0}
\definecolor{nvteal}{HTML}{00A5A5}
\definecolor{nvlid}{HTML}{C38FE3}

\definecolor{granaryblue}{HTML}{4E7A00}
\definecolor{granarylight}{HTML}{E8F5CC}
\definecolor{granarygray}{HTML}{F4F4F2}
\definecolor{granarylink}{HTML}{76B900}
\input{figures/figure_colors}
\newcolumntype{Y}{>{\raggedright\arraybackslash}X}
\newcommand{\transcribename}{Nemotron~3.5 Transcribe}
\lstdefinestyle{granaryprompt}{
  basicstyle=\ttfamily\footnotesize,
  breaklines=true,
  breakatwhitespace=false,
  columns=fullflexible,
  keepspaces=true,
  showstringspaces=false,
  frame=single,
  rulecolor=\color{nvgrid},
  backgroundcolor=\color{nvpanel},
  xleftmargin=5pt,
  xrightmargin=5pt,
  framexleftmargin=4pt,
  framexrightmargin=4pt,
  aboveskip=8pt,
  belowskip=8pt
}

\title{\vspace{-1em}{\color{nvgreen}\rule{\linewidth}{2pt}}\\[0.3em]
  \textbf{Granary-v2: Scalable Multilingual Speech Data\\
  Curation for Predictable and Controllable Model Outputs}\\[0.2em]
  \large Technical Report\\[-0.2em]
  {\color{nvgreen}\rule{\linewidth}{2pt}}}
\author{NVIDIA\\[0.4em]
  {\small\href{https://huggingface.co/datasets/nvidia/Granary-v2}{\textcolor{nvgreen}{Dataset: nvidia/Granary-v2}}}}
\date{}

\hypersetup{
  pdftitle={Granary-v2: Scalable Multilingual Speech Data Curation for Predictable and Controllable Model Outputs},
  pdfauthor={NVIDIA},
  pdfsubject={Multilingual speech data curation and labeling},
  pdfkeywords={speech recognition, multilingual speech, data curation, text normalization, NeMo Curator}
}

\makeatletter
\renewcommand{\maketitle}{%
  \begingroup
    \centering
    {\fontsize{16}{19}\selectfont\@title\par}%
    \vspace{0.4em}%
    {\large\@author\par}%
  \endgroup
  \vspace{0.2em}%
}
\makeatother
\renewenvironment{abstract}{%
  \small
  \begin{center}\bfseries\abstractname\end{center}%
  \begin{list}{}{\setlength{\leftmargin}{12pt}\setlength{\rightmargin}{12pt}}%
  \item\relax
}{\end{list}}

\begin{document}
\raggedbottom
\maketitle
\pagestyle{plain}

\input{sections/abstract}
\input{figures/curation_summary}

\clearpage
\tableofcontents
\clearpage
\input{sections/introduction}
\input{sections/related_work}
\input{sections/curation_system}
\input{sections/pipeline_application}
\input{sections/dataset}
\input{sections/evaluation}
\input{sections/responsible_release}
\input{sections/conclusion}

\FloatBarrier
\input{sections/contributors}

\bibliographystyle{plainnat}
\bibliography{references}

\clearpage
\appendix
\input{sections/appendices}

\end{document}

%% file: figures/figure_colors.tex
\definecolor{graybg}{HTML}{E9EDF2}
\definecolor{graybdr}{HTML}{627282}
\definecolor{bluebg}{HTML}{CCE3F6}
\definecolor{bluebdr}{HTML}{1B70A8}
\definecolor{purplebg}{HTML}{E5D5EF}
\definecolor{purplebdr}{HTML}{864CB5}
\definecolor{orangebg}{HTML}{FFE1B5}
\definecolor{orangebdr}{HTML}{C56A12}
\definecolor{greenbg}{HTML}{D6E9B7}
\definecolor{greenbdr}{HTML}{598B14}

%% file: sections/abstract.tex
\begin{abstract}
Speech-language models that follow instructions need training data that connects the same recording to different output forms and tasks. We present Granary-v2, a multilingual speech dataset and reusable curation system that provides a data foundation for these models. Granary-v2 extends Granary-v1 with aligned transcript variants and instruction--target pairs for multi-task speech-language training. Language coverage increases from 25 to 48 languages. Built with NeMo Curator, the system combines language-dependent ASR labeling, transcript improvement, and text label generation. It produces aligned targets for speech-faithful transcription, written forms with and without disfluencies, and code-switched transcription. Additional targets include transcript-derived captions and question--answer pairs, with context and translations for selected records. The full processed collection contains approximately 843,600 unique audio hours, with translations covering about 142,800 hours into English and about 55k hours of English audio into 31 target languages. We evaluate label-quality filtering with two English Transformer-TDT models under the same training-step budget. Filtering reduces the unweighted mean word error rate across nine benchmarks from 4.91\% to 4.67\%, with improvements on six benchmarks. We also report curation throughput and downstream ASR and speech translation results from Nemotron-3.5-Transcribe, which uses Granary-v2 as part of its training data. We release the Granary-v2 pipeline\footnote{\url{https://github.com/NVIDIA-NeMo/Curator/tree/main/tutorials/audio/granary_v2}}, built with NeMo Curator, and model training recipes\footnote{\url{https://github.com/NVIDIA-NeMo/Speech}} as open source. We release a public subset of the generated labels and instruction--target pairs through Hugging Face\footnote{\url{https://huggingface.co/datasets/nvidia/Granary-v2}}.
\end{abstract}

%% file: figures/curation_summary.tex
\begingroup
\setlength{\intextsep}{6pt}
\begin{figure}[H]
\centering
\includegraphics[width=0.90\linewidth]{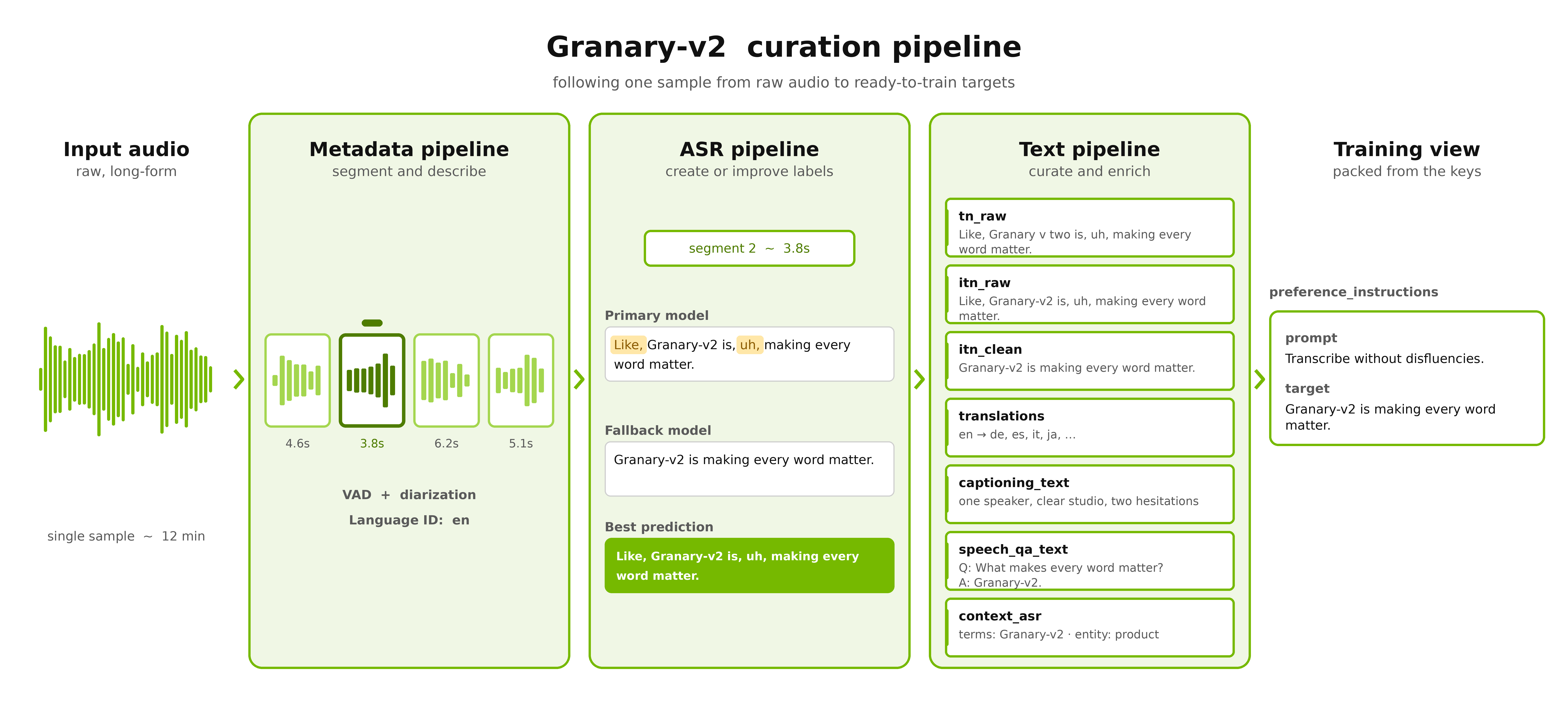}
\caption{Illustrative Granary-v2 curation flow and instruction--target packing.}
\label{fig:curation-summary}
\end{figure}
\endgroup

%% file: sections/introduction.tex
\section{Introduction}
\label{sec:introduction}

Automatic speech recognition (ASR) serves applications with different output requirements. A dictation system may need numbers written as digits and hesitation markers removed. A transcript for linguistic analysis may need spoken number forms, repetitions, and false starts. Other applications need domain context, code-switched text, or translation. Speech-language models can express these requirements through natural language prompts, but training them requires examples that connect each instruction to the appropriate output. Training for these capabilities requires consistent targets derived from the same speech, together with instructions that state the required task and output form. But most of speech corpora typically provide one transcript for each recording, and their transcription conventions differ. A single reference does not show a model how to produce several valid output forms or how to select among them. 

We introduce Granary-v2, a multilingual speech dataset and reusable curation system that provides a data foundation for speech-language models. Granary-v2 supports two training stages: speech-encoder pretraining and multi-task training with instruction--target pairs. These pairs cover transcription conventions, disfluency control, contextual ASR, code-switched transcription, transcript-derived captions and question answering, and speech translation. 
Granary-v2 builds these targets around a common \emph{speech-faithful transcript}. This representation preserves spoken content, including spoken number forms, grammar, fillers, repetitions, and false starts. The text pipeline derives written forms and other task targets from this common source. Instructions specify which target a model must produce. This design supports training for prompt-controlled outputs while maintaining a common source for the labels.
Three separate pipelines perform metadata extraction, ASR labeling, and text label generation. Built with NeMo Curator, the system supports language-specific labeling routes, quality filtering, and restartable processing. This report describes the system, the released data, and evaluations of label filtering and downstream use.

We make the following contributions:
\begin{enumerate}
  \item A multilingual curation system built with NeMo Curator, with separate metadata, ASR, and text pipelines.
  \item Aligned text targets for spoken forms, written forms, code-switched text, and disfluency handling, with optional context and translation labels.
  \item Processing methods that record label sources and quality decisions and support recovery after job failure.
  \item A processed collection with 843,579.22 unique ASR hours across 48 languages and 19 corpus entries, with 142,778.43 hours of translation into English. The public subset excludes labels and instruction pairs from YODAS2, YouTube-Commons, and ReazonSpeech-v2~\citep{granary-v2-data}.
  \item An English ASR comparison of high-quality and all labels, measured curation throughput, and qualitative text-model comparisons.
\end{enumerate}
We evaluate quality filtering by training English ASR on full dataset  and on the high-quality subset under the same training budget. Section~\ref{sec:evaluation} presents this comparison and describes downstream ASR and speech translation results from a speech-language model trained with Granary-v2.

%% file: sections/related_work.tex
\section{Related Work}
\label{sec:related-work}
We compare Granary-v2 with large speech corpora, automated curation systems, and datasets with several text targets per utterance in Table~\ref{tab:dataset-comparison}. 

\subsection{Large and Multilingual Speech Datasets}
\input{tables/dataset_comparison}

Granary-v1 combines YODAS, YouTube-Commons, and MOSEL through segmentation, ASR, hallucination filtering, punctuation restoration, and translation~\citep{granary-v1,youtube-commons}. Granary-v2 extends this approach with several text targets for the same audio. These include spoken forms and written forms with and without disfluencies. Selected utterances also receive context and translation labels.

YODAS collects online video audio with uploaded or automatic captions where available~\citep{yodas}. YODAS v3 contains over 1.1 million hours of 48~kHz multichannel audio in more than 100 languages~\citep{yodas-v3}. It provides captions and English translations where available. MOSEL combines European speech sources and uses automatic transcription~\citep{mosel}. GigaSpeech~2 uses iterative pseudo-labeling and filtering for Thai, Indonesian, and Vietnamese~\citep{gigaspeech2}. Emilia uses segmentation, transcription, and filtering for speech generation data~\citep{emilia}.
MLS aligns audiobook recordings with book text~\citep{mls}. VoxPopuli includes parliamentary speech, transcriptions, and aligned interpretations~\citep{voxpopuli}. Common Voice collects  recordings of prompted speech and uses community validation~\citep{commonvoice,commonvoice-v26}.

\subsection{Automated Speech Data Curation}

Automated curation commonly combines speech segmentation, speaker analysis, ASR, and filtering. AutoPrep uses speech enhancement, VAD, speaker clustering, target-speaker extraction, and multilingual ASR~\citep{autoprep}. Emilia-Pipe uses source separation, diarization, transcription, and filtering for speech generation~\citep{emilia}.

Granary-v2 supports both audio without trusted text and recordings with existing human transcripts. It adds language-dependent model selection and retains intermediate predictions and quality decisions in internal processing manifests. Released records expose the generated targets and the high-quality selection flag. Section~\ref{sec:system} describes these methods.

\subsection{Speech-Language and Contextual-ASR Data}

Transcription conventions affect how models handle spontaneous speech. Fillers, repetitions, and self-corrections can be retained or removed. Work on controllable verbatim ASR studies different transcript styles and output selection~\citep{transcription-policy}. Preference-ASR evaluates instruction control of number forms, entities, disfluencies, and case~\citep{koluguri2026preferenceasr}.

Contextual ASR uses names, terms, or supporting material to guide recognition. ConEC pairs earnings-call recordings with slides, news releases, and participant names~\citep{conec}. CoVoST~2 pairs speech with transcripts and translations~\citep{covost2}. SALM studies speech recognition and translation with in-context learning~\citep{chen2023salm}. SpeechVerse uses audio--instruction--response examples for ASR, translation, and speech understanding~\citep{speechverse}.

Granary-v2 associates several aligned text targets with the same audio. These targets support multi-task training for transcription, output formatting, contextual ASR, question answering, and translation.

%% file: tables/dataset_comparison.tex
\begin{table}[t]
\centering
\caption{
Comparison of Granary-v2 with large-scale speech datasets. Granary-v2 statistics describe the full processed collection, including YODAS2, YouTube-Commons, and ReazonSpeech-v2. Labels from these three sources are excluded from the public release.
}
\label{tab:dataset-comparison}

\small
\setlength{\tabcolsep}{4pt}
\renewcommand{\arraystretch}{1.25}

\begin{tabularx}{\linewidth}{@{}
    >{\raggedright\arraybackslash}p{0.36\linewidth}
    >{\centering\arraybackslash}p{0.14\linewidth}
    >{\centering\arraybackslash}p{0.10\linewidth}
    Y
@{}}
\toprule
\textbf{Dataset} &
\textbf{Hours} &
\textbf{Langs} &
\textbf{Text Supervision} \\
\midrule

\textbf{Granary-v2 (processed)}~\citep{granary-v2-data} &
\textbf{843,579} &
\textbf{48} &
Multiple aligned transcript forms; contextual ASR and translation \\

Granary-v1~\citep{granary-v1} &
643,000 &
25 &
ASR transcripts; speech translation \\

MOSEL~\citep{mosel} &
441,000 &
24 &
ASR transcripts \\

YODAS~\citep{yodas} &
$\sim$500,000 &
100+ &
Captions (where available) \\

YODAS v3~\citep{yodas-v3} &
$>1{,}100{,}000$ &
100+ &
Captions; English translations (where available) \\

GigaSpeech 2~\citep{gigaspeech2} &
$\sim$30,000 &
3 &
ASR transcripts \\

Emilia~\citep{emilia} &
101,654 &
6 &
ASR transcripts \\

Common Voice v26~\citep{commonvoice-v26} &
28,893 &
294 &
ASR transcripts \\

VoxPopuli~\citep{voxpopuli} &
1,800 &
23 &
ASR transcripts \\

MLS~\citep{mls} &
$\sim$50,000 &
8 &
ASR transcripts \\

\bottomrule
\end{tabularx}
\end{table}

%% file: sections/curation_system.tex
\section{The Granary-v2 Curation System}
\label{sec:system}

\input{figures/pipeline_overview}

We build the curation system with NeMo Curator~\citep{nemo-curator}. It accepts audio with or without existing transcripts. Figure~\ref{fig:pipeline-overview} shows how three separate pipelines prepare audio, produce transcripts, and derive training targets.

The metadata pipeline divides raw or long recordings into short speech segments. It adds information about speakers, language, and sound events. The ASR pipeline then generates or improves transcripts using language-specific models and validation checks. The text pipeline derives different transcript forms and task labels from these transcripts. It applies quality checks, adds context and translations for selected records, and assembles instruction--target pairs. These pairs connect each recording to instructions that specify the task and output form.

Each pipeline can run independently. Already segmented audio can enter the ASR pipeline directly. This lets us select the processing steps for the available inputs and required training targets.

The released records contain audio identifiers, source corpus and language identifiers, text targets, optional context and translations, instruction--target pairs, and a \texttt{high\_quality} flag. Internal processing manifests also retain intermediate model predictions and stage decisions. Appendix~\ref{app:quality-rules} describes the released fields. The following subsections describe each pipeline in detail.

\subsection{Metadata Extraction Pipeline}
\label{sec:metadata}

\input{figures/metadata_pipeline}

The metadata pipeline converts raw or long recordings into short speech segments. It accepts manifests that reference separate audio files or tarred audio shards. Short recordings can bypass this pipeline or use it to add metadata.

We decode each recording, convert it to mono, and resample it to 16~kHz. The \path|nvidia/Nemotron-3-Diarization| checkpoint processes the full recording~\citep{streaming-sortformer,nemotron3-diarization}. It supports up to eight speakers and records speaker activity before segmentation. Appendix~\ref{app:reference-config} describes the reference settings. The speaker timeline is saved in RTTM format and retained for later use.

Silero voice activity detection (VAD) divides the recording into speech segments between 0.5 and 40 seconds~\citep{silero-vad}. We match each segment to the speaker timeline to derive its speaker count. PANNs CNN14 identifies speech, music, and other sound events in each segment~\citep{panns}. SpeechBrain's VoxLingua107 ECAPA-TDNN model predicts the spoken language and a confidence value between zero and one~\citep{voxlingua107,speechbrain-voxlingua107-ecapa}.

The writer produces 16~kHz mono Opus clips and a JSONL manifest. Each entry includes duration, source recording, segment interval, source audio format, speaker count, language prediction, and sound events. Existing transcripts and source identifiers are retained where applicable.

Inputs with no detected speech receive a \texttt{vad\_empty} entry. Inputs that cannot be decoded receive a \texttt{read\_error} entry. A completion marker is written after every unique input in a shard has been accounted for. This permits completeness checks before a job resumes. The measured rate is about 23 audio-hours per GPU-hour for the documented workload.

\subsection{ASR Labeling Pipeline}
\label{sec:asr-pipeline}

\input{figures/asr_pipeline}
\input{tables/asr_modes}

The ASR pipeline has two modes. Audio without a trusted reference uses primary and recovery ASR models. Audio with a source transcript uses that transcript as an additional reference. A language-dependent router selects the models. Table~\ref{tab:asr-groups} gives the routing groups. Appendix~\ref{sec:multilingual-routing} lists their languages and model names.

For audio-only Qwen3-Omni labeling~\citep{xu2025qwen3omni}, the prompt requests fillers, repetitions, false starts, colloquial expressions, and spoken number forms. An available source transcript can supply trusted spellings. Appendix~\ref{app:reference-prompts} lists the reference-improvement prompts. For languages outside the supported improvement path, the pipeline retains the aligned source label.

The selection stage checks the primary and recovery predictions for known hallucination patterns. It selects the primary prediction by default and uses the recovery prediction when the primary fails validation. If both trigger a hallucination rule but have zero pairwise word or character error rate, the agreement exception retains the primary.

Regular-expression cleaning removes invalid symbols, standardizes punctuation, and collapses excess whitespace. A language-dependent stage joins isolated uppercase letters in abbreviations, such as ``A P I'' to ``API''.

A record with no valid candidate, or with empty text after cleaning, remains in the processing manifest. The pipeline records the reason and excludes the record from later stages that need a valid transcript.

\begin{table}[t]
\centering
\small
\caption{Language-dependent ASR model routing. Appendix~\ref{sec:multilingual-routing} lists checkpoint identifiers and language membership.}
\label{tab:asr-groups}
\begin{tabularx}{\linewidth}{@{}cYYr@{}}
\toprule
Group & Primary model & Recovery model & Languages \\
\midrule
A & Qwen3-Omni & Qwen3-ASR~\citep{shi2026qwen3asr} & 15 \\
B & Qwen3-ASR & Whisper~\citep{radford2022whisper,whisper-large-v3} & 3 \\
C & Parakeet v3~\citep{sekoyan2025canaryv2,parakeet-v3} & Whisper & 8 \\
D & Whisper & Parakeet v3 & 9 \\
E & Hebrew Whisper~\citep{hebrew-whisper} & Whisper & 1 \\
F & Indic-Transcribe-core & IndicConformer & 11 \\
G & Indic-Transcribe-core & Internal Hindi Parakeet & 1 \\
\midrule
\multicolumn{3}{@{}l}{Total configured languages} & 48 \\
\bottomrule
\end{tabularx}
\end{table}

The Indic primary model is \path|bodhan-ai/indic-transcribe-core|~\citep{indic-transcribe-core}. Group F uses monolingual IndicConformer recovery models~\citep{indicconformer-monolingual}. The router selects a separate checkpoint for each language. Hindi uses Internal Hindi Parakeet as its recovery model.

\subsection{Text Label Generation Pipeline}
\label{sec:text-pipeline}

\input{figures/text_pipeline}

The text pipeline reads the cleaned ASR transcript and produces separate targets. Each task prompt defines a text transformation, such as punctuation, number conversion, or disfluency removal.

We use Gemma-4-31B-it-FP8-dynamic for Indic languages~\citep{gemma4-31b,gemma4-fp8}. All other languages use Qwen3.6-27B-FP8~\citep{qwen36-27b}. We selected the latter based on qualitative observations of word preservation and on processing cost. Appendix~\ref{app:text-model-comparison} presents the comparison.

\subsubsection{Core Text Stages}

\input{tables/output_targets}

\paragraph{Entity recovery.}
For records with a trusted source transcript, we restore the spelling and capitalization of letter-only names and acronyms. The stage compares the ASR transcript with the source annotation. It is skipped when no source reference is available or the selected transcript already matches that reference.

\paragraph{Punctuation and capitalization.}
The general prompt permits changes to punctuation and case while retaining the words in their original order. Output is stored in \texttt{pnc\_text}. Validation checks can restore the input text after a failed check. Appendix~\ref{app:validation-config} describes the checks and fallback rules.

\paragraph{Text normalization.}
Text normalization expands numbers, dates, times, currencies, units, percentages, and symbols into spoken forms. The prompt retains fillers, repetitions, false starts, grammar, and code-switched content. The model infers the spoken form from the transcript and stores it in \texttt{tn\_raw}.

\paragraph{Transcript-language identification.}
The language model reads \texttt{pnc\_text} and returns the primary language name and a list of languages present. A verification step resolves names and codes before comparison with the source language. This check differs from acoustic language identification. A mismatch excludes the record from later stages unless a configured language-pair exception applies.

\paragraph{Inverse text normalization.}
We derive two written forms from \texttt{tn\_raw}. Standard inverse text normalization (ITN) restores written numbers, dates, currencies, units, and symbols while preserving disfluencies. Its output is \texttt{itn\_raw}. A separate stage reads \texttt{itn\_raw}, removes disfluencies, and writes \texttt{itn\_clean}. The serial reference path checks word counts and can restore the input written form after excessive deletion. Appendix~\ref{app:validation-config} gives the word-count threshold and fallback rule.

\paragraph{Code-switched text.}
Code-switching restoration reads \texttt{tn\_raw} and writes a separate \path|code_switched_text| target. It restores standard Latin spellings of English-origin words represented phonetically in another script. Native-language words retain their script. The prompt preserves word order, punctuation, disfluencies, and native-script suffixes attached to English roots. It leaves uncertain words unchanged.

\paragraph{Captioning and speech question answering.}
Both stages read \texttt{tn\_raw}. Caption generation writes a short summary to \texttt{captioning\_text}, with a prompt that requests the transcript language. Speech question answering produces at most one question--answer pair supported by the transcript. It stores the result in \texttt{speech\_qa\_text}, or returns \texttt{SKIP} when the text is unsuitable. Both tasks use transcript content as their source.

\subsubsection{Quality Filtering}

Quality filtering runs on CPUs and selects records using structural and transcript-agreement checks.

Quality checks cover incorrect abbreviation joins, script-dependent filtering errors, character loss, and language mismatches. These checks identify errors that can change labels or exclude valid records without causing a processing failure.

\paragraph{Structural checks.}
We check character rate, known hallucination phrases at the start of the text, and residual ASCII digits in the spoken-form field. The default character-rate interval is 4--30 characters per second. Configured exceptions reduce the floor to 1.5 for Chinese, Japanese, Korean, and Cantonese, and to 2.0 for Dusha and Golos. The ceiling remains 30. We use these empirical thresholds to identify unusual text-to-duration ratios. The digit rule matches the ASCII range \texttt{[0-9]}.

\paragraph{Agreement scoring.}
For newly generated labels, we compare the primary transcript as reference with the recovery transcript as hypothesis. For label improvement, we compare the source human transcript as reference with the primary prediction. We avoid comparing a selected copy of the human transcript with itself.

English uses the Open ASR Leaderboard English normalizer. Other languages use its basic multilingual normalizer~\citep{srivastav2025openasr}. The same normalizer is applied to both texts. Word error rate (WER) is used for word-scored languages. Character error rate (CER) is used for the configured character-scored languages. The CER set differs from the character-rate exceptions. Appendix~\ref{app:quality-rules} gives the configured sets and human-reference exceptions.

\begin{table}[t]
\centering
\small
\caption{Base agreement rules after structural checks. Human-reference exceptions are described in Appendix~\ref{app:quality-rules}.}
\label{tab:hq-thresholds}
\begin{tabularx}{\linewidth}{@{}p{0.35\linewidth}Yl@{}}
\toprule
Case & Condition & Decision \\
\midrule
Character-scored language & $\mathrm{CER}\leq0.05$ & Accept \\
Word-scored, ITN unchanged & $\mathrm{WER}\leq0.01$ & Accept \\
Word-scored, ITN modified & $\mathrm{WER}-\overline{\mathrm{WER}}_{\mathrm{group}}\leq0.01$ & Accept \\
Other cases & No applicable acceptance condition & Reject \\
\bottomrule
\end{tabularx}
\end{table}

For modified ITN, the group is the corpus--language pair. Its mean excludes skipped records and non-finite scores. We use agreement between predictions, or agreement with a source reference, as a consistency score for label selection.

\subsubsection{Contextual-ASR Labels}

Context generation runs on records that pass quality filtering. Qwen3.6-27B-FP8 extracts the domain, entities, entity categories, and context terms~\citep{qwen36-27b}. It also generates plausible but irrelevant terms as semantic distractors. Appendix~\ref{app:context-prompts} gives the generation prompt. Appendix~\ref{app:packing-details} gives the related training instructions, sampling rules, and phonetic-matching settings. Where a phoneme vocabulary is available, phonetic matching adds words that sound similar to the relevant terms.

\input{tables/context_variants}

The stage stores terms, categories, distractors, and five prompt variants in \texttt{context\_asr}. Table~\ref{tab:context-variants} describes the supplied context. Each variant retains the transcript target for the same audio, including when hints are incomplete or irrelevant. The measured rate is about 30 audio-hours per node-hour for the selected high-quality records.

\subsubsection{Translation}
\label{sec:translation}

We translate speech-faithful normalized text using Qwen3.5-122B-A10B-FP8~\citep{qwen35}. Early translation runs used the full data rather than only the high-quality subset. Later runs mainly used high-quality records. The release includes translations from both, so translation coverage is not restricted to records with the final high-quality flag.

In experiments on FLEURS~\citep{conneau2022fleurs}, this model achieved higher COMET scores~\citep{comet} than the 27B and 35B variants, particularly for translation from English into other languages. Translation quality further improved when the prompt included a system instruction and the full names of the source and target languages.

The prompt instructs the model to return only the translation, without explanations or additional content, and to preserve incomplete utterances present in speech-derived text. English records are translated into 31 target languages, whereas non-English records are translated into English. To reduce the computational cost of translating English into many languages, we optionally translate a 10\% sample of the eligible English records. The system and user prompts are provided in Appendix~\ref{app:translation-prompts}.

We discard translations that are empty or exceed the length limits. For each remaining translation, we use Cometoid22-WMT23~\citep{gowda-etal-2023-cometoid} (\href{https://huggingface.co/marian-nmt/cometoid22-wmt23}{\path|marian-nmt/cometoid22-wmt23|}) through PyMarian to assign a reference-free quality-estimation score. We store this score alongside the translation, enabling practitioners to filter the data by increasingly strict quality thresholds during training.

\subsubsection{Instruction Packing}

A deterministic CPU stage assembles instruction--target pairs from existing text targets, context prompts, and valid translations. Each entry in \texttt{preference\_instructions} contains an instruction, target text, and tags for task and language. These pairs associate the same audio with several training targets.

\input{tables/instruction_targets}

Table~\ref{tab:instruction-targets} maps instruction families to target fields. The reference packer prefers spoken-form text for transcription and uses a configured fallback when that field is absent. A translation-score threshold selects training pairs after generation, while the translation manifest retains the scored outputs. Appendix~\ref{app:packing-details} describes missing values, sampling, and selected instructions.

\subsection{Example Records and Training Targets}
\label{sec:output-example}

\input{figures/output_example}

Pairing each representation with a matching instruction lets a speech-language model learn output selection from the same audio, as illustrated in Figure~\ref{fig:output-example}.

\subsection{Reliability and Scale}

Production runs use sharded inputs, model-server reuse, completion markers, output-count checks, and structured status files. Shard markers avoid repeating completed shards. Field checks avoid repeating completed stages within a record. We check expected row counts before treating a marker as valid.

We divide work mainly by audio duration because shard and utterance counts can hide large differences in workload. Job size must account for model loading, initialization, and the available time limit. Utterance length, per-record overhead, and optional-stage eligibility also affect throughput.

Model servers use NVIDIA Dynamo and vLLM~\citep{dynamo,vllm}. Scaling combines SLURM arrays, a streaming Ray graph, stage workers, GPU placement, model-specific batching, and concurrent storage I/O~\citep{slurm,ray}. Separate jobs with balanced durations reduced shared-writer and object-store contention in the documented ASR workloads.

\input{tables/throughput}

The production measurements use A100 80~GB GPUs. The GPU-node rates assume eight allocated GPUs. Metadata is measured per GPU, while quality filtering runs on CPUs. ASR labeling processes about 281 audio-hours per node-hour on the short-utterance workload.

Text-enrichment rates range from 15 to 25 audio-hours per node-hour across workloads and job sizes. Context timing includes reading, the quality gate, context generation, prompt variants, and writing. Translation timing covers eligible Japanese audio translated into English.

For a node-based rate $R$, processing $H$ audio hours takes an estimated $H/R$ node-hours. With $g$ allocated GPUs, the corresponding estimate is $gH/R$ GPU-hours. Logged compute uses allocation times and includes failures and retries. Appendix~\ref{app:compute-method} defines the accounting method.

%% file: figures/pipeline_overview.tex
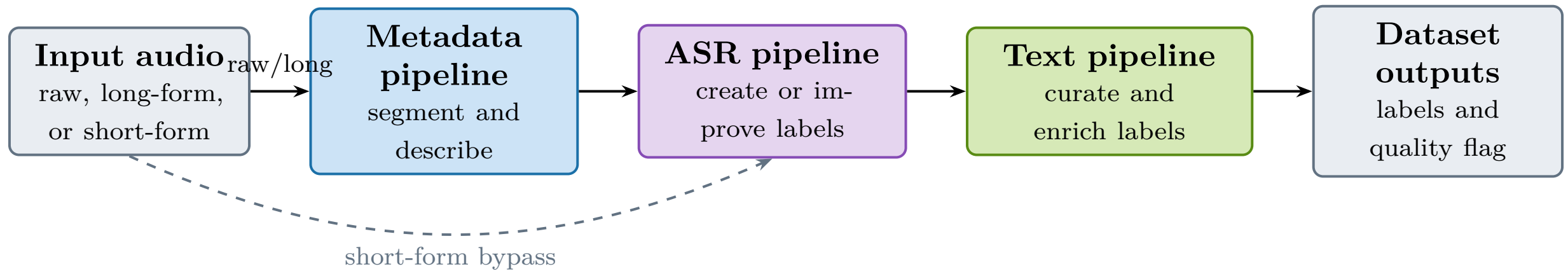
\begin{figure}[htbp]
\centering

\resizebox{0.98\linewidth}{!}{%
\begin{tikzpicture}[
  font=\small,
  box/.style={
    draw,
    rounded corners=3pt,
    thick,
    inner sep=5pt,
    align=center,
    minimum height=0.95cm
  },
  arw/.style={
    -{Stealth[length=5pt]},
    thick
  },
  skip/.style={
    -{Stealth[length=5pt]},
    thick,
    dashed,
    graybdr
  },
]

  \node[
    box,
    text width=2.3cm,
    fill=graybg,
    draw=graybdr
  ] (input) {
    \textbf{Input audio}\\[-1pt]
    {\scriptsize raw, long-form, or short-form}
  };

  \node[
    box,
    text width=2.6cm,
    fill=bluebg,
    draw=bluebdr,
    right=0.65cm of input
  ] (metadata) {
    \textbf{Metadata pipeline}\\[-1pt]
    {\scriptsize segment and describe}
  };

  \node[
    box,
    text width=2.6cm,
    fill=purplebg,
    draw=purplebdr,
    right=0.65cm of metadata
  ] (asr) {
    \textbf{ASR pipeline}\\[-1pt]
    {\scriptsize create or improve labels}
  };

  \node[
    box,
    text width=2.8cm,
    fill=greenbg,
    draw=greenbdr,
    right=0.65cm of asr
  ] (text) {
    \textbf{Text pipeline}\\[-1pt]
    {\scriptsize curate and enrich labels}
  };

  \node[
    box,
    text width=2.4cm,
    fill=graybg,
    draw=graybdr,
    right=0.65cm of text
  ] (output) {
    \textbf{Dataset outputs}\\[-1pt]
    {\scriptsize labels and quality flag}
  };

  \draw[arw]
    (input) -- node[above, font=\scriptsize] {raw/long} (metadata);

  \draw[arw]
    (metadata) -- (asr);

  \draw[arw]
    (asr) -- (text);

  \draw[arw]
    (text) -- (output);

  \draw[skip]
    (input.south)
    to[bend right=24]
    node[below, font=\scriptsize, text=graybdr]
    {short-form bypass}
    (asr.south);

\end{tikzpicture}%
}

\caption{Top-level Granary-v2 curation flow. The three pipelines are reusable and independently selectable.}
\label{fig:pipeline-overview}
\end{figure}

%% file: figures/metadata_pipeline.tex
\begin{figure}[b]
\centering

\resizebox{\linewidth}{!}{%
\begin{tikzpicture}[
  font=\small,
  box/.style={
    draw,
    rounded corners=3pt,
    thick,
    inner sep=5pt,
    text width=#1,
    align=center,
    minimum height=1.35cm
  },
  box/.default=2.2cm,
  arw/.style={
    -{Stealth[length=5pt]},
    thick
  },
]

  \node[
    box=2.2cm,
    fill=graybg,
    draw=graybdr
  ] (input) {
    \textbf{Input recording}\\[1pt]
    {\scriptsize raw or long-form audio}
  };

  \node[
    box=2.5cm,
    fill=bluebg,
    draw=bluebdr,
    right=0.6cm of input
  ] (diar) {
    \textbf{Speaker}\\[-1pt]
    \textbf{diarization}\\[1pt]
    {\scriptsize Nemotron 3 Diarization}\\
    {\scriptsize on the complete recording}
  };

  \node[
    box=2.4cm,
    fill=orangebg,
    draw=orangebdr,
    right=0.6cm of diar
  ] (vad) {
    \textbf{Speech}\\[-1pt]
    \textbf{segmentation}\\[1pt]
    {\scriptsize VAD fan-out}\\
    {\scriptsize 0.5--40\,s segments}
  };

  \node[
    box=2.5cm,
    fill=purplebg,
    draw=purplebdr,
    right=0.6cm of vad
  ] (sed) {
    \textbf{Sound-event}\\[-1pt]
    \textbf{detection}\\[1pt]
    {\scriptsize speech, music, and}\\
    {\scriptsize other events}
  };

  \node[
    box=2.5cm,
    fill=purplebg,
    draw=purplebdr,
    right=0.6cm of sed
  ] (lid) {
    \textbf{Language}\\[-1pt]
    \textbf{identification}\\[1pt]
    {\scriptsize SpeechBrain}
  };

  \node[
    box=2.6cm,
    fill=greenbg,
    draw=greenbdr,
    right=0.6cm of lid
  ] (output) {
    \textbf{Segment outputs}\\[1pt]
    {\scriptsize Opus clips and}\\
    {\scriptsize JSONL metadata}
  };

  \draw[arw] (input) -- (diar);
  \draw[arw] (diar) -- (vad);
  \draw[arw] (vad) -- (sed);
  \draw[arw] (sed) -- (lid);
  \draw[arw] (lid) -- (output);

\end{tikzpicture}%
}

\caption{Metadata extraction. Nemotron 3 Diarization processes the full recording and saves a speaker timeline in RTTM format. VAD divides the recording into speech segments. We derive each segment's speaker count from that timeline. Sound-event detection and language identification run on the segments. Outputs include 16~kHz mono Opus clips, a JSONL manifest, and the retained RTTM files.}
\label{fig:metadata-pipeline}
\end{figure}
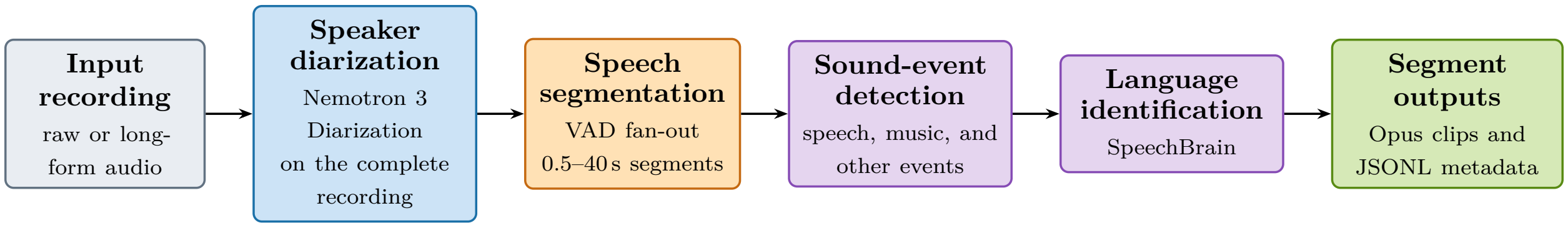

%% file: figures/asr_pipeline.tex
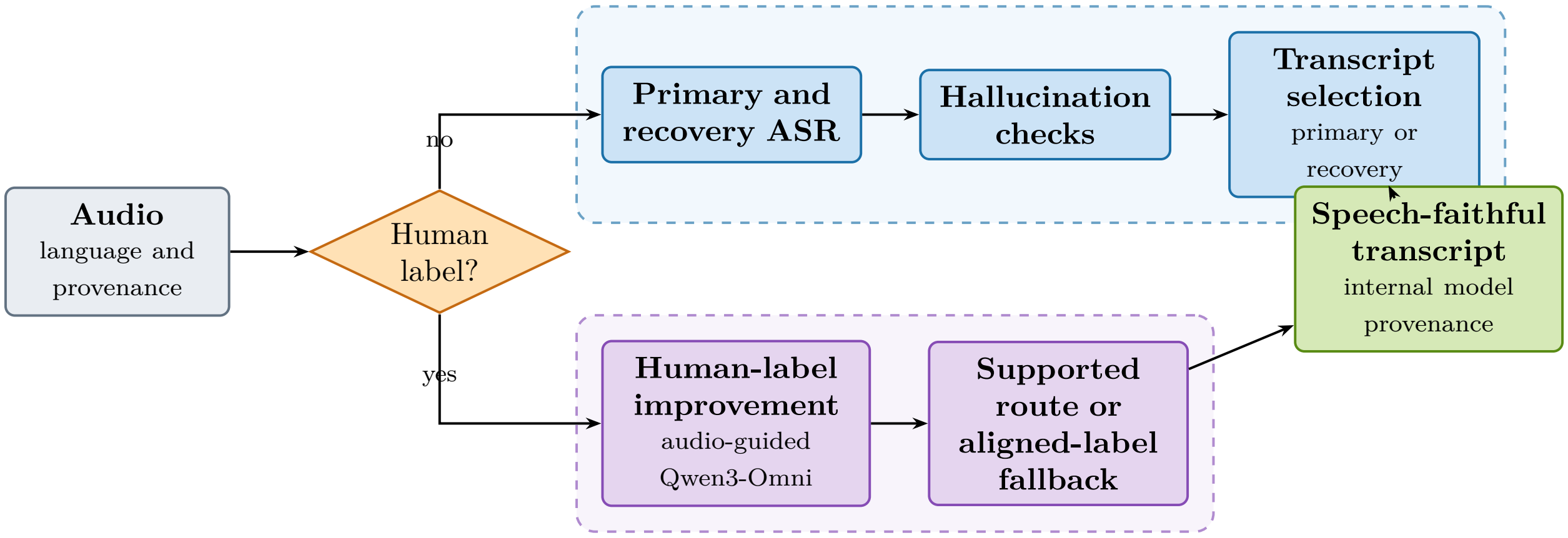
\begin{figure}[htbp]
\centering
\resizebox{\textwidth}{!}{%
\begin{tikzpicture}[
  font=\small,
  box/.style={draw, rounded corners=3pt, thick, inner sep=5pt,
              text width=#1, align=center, minimum height=1.0cm},
  box/.default=2.2cm,
  decision/.style={draw=orangebdr, diamond, thick, fill=orangebg,
                   aspect=2.1, align=center, inner sep=2pt},
  arw/.style={-{Stealth[length=5pt]}, thick},
  route/.style={draw, rounded corners=6pt, thick, dashed, inner sep=8pt},
]
  \node[box=2.2cm, fill=graybg, draw=graybdr] (input)
    {\textbf{Audio}\\[-1pt]{\scriptsize language and provenance}};
  \node[decision, right=0.9cm of input] (human) {Human\\label?};

  \node[box=2.6cm, fill=bluebg, draw=bluebdr,
        above right=0.65cm and 1.1cm of human] (primary)
    {\textbf{Primary and recovery ASR}};
  \node[box=2.5cm, fill=bluebg, draw=bluebdr, right=0.65cm of primary] (check)
    {\textbf{Hallucination checks}};
  \node[box=2.5cm, fill=bluebg, draw=bluebdr, right=0.65cm of check] (recovery)
    {\textbf{Transcript selection}\\[-1pt]{\scriptsize primary or recovery}};

  \node[box=2.7cm, fill=purplebg, draw=purplebdr,
        below right=0.65cm and 1.1cm of human] (improve)
    {\textbf{Human-label improvement}\\[-1pt]{\scriptsize audio-guided Qwen3-Omni}};
  \node[box=2.6cm, fill=purplebg, draw=purplebdr, right=0.65cm of improve] (fallback)
    {\textbf{Supported route or aligned-label fallback}};

  \node[box=2.7cm, fill=greenbg, draw=greenbdr,
        right=1.0cm of $(recovery)!0.5!(fallback)$] (output)
    {\textbf{Speech-faithful transcript}\\[-1pt]{\scriptsize internal model provenance}};

  \draw[arw] (input) -- (human);
  \draw[arw] (human) |- node[pos=0.2, above, font=\scriptsize] {no} (primary);
  \draw[arw] (primary) -- (check);
  \draw[arw] (check) -- (recovery);
  \draw[arw] (recovery) -- (output);
  \draw[arw] (human) |- node[pos=0.2, below, font=\scriptsize] {yes} (improve);
  \draw[arw] (improve) -- (fallback);
  \draw[arw] (fallback) -- (output);

  \begin{scope}[on background layer]
    \node[route, draw=bluebdr!65, fill=bluebg!25,
          fit=(primary)(check)(recovery)] {};
    \node[route, draw=purplebdr!65, fill=purplebg!25,
          fit=(improve)(fallback)] {};
  \end{scope}
\end{tikzpicture}%
}
\caption{ASR-labeling pipeline. Inputs without human transcripts use language-dependent primary and recovery ASR. Inputs with human transcripts use audio-guided Qwen3-Omni improvement when supported and retain aligned labels otherwise.}
\label{fig:asr-pipeline}
\end{figure}

%% file: tables/asr_modes.tex
\begin{table}[H]
\centering
  \caption{The two ASR-labeling modes.}
  \label{tab:asr-modes}
  \centering
  \small
  \begin{tabularx}{\linewidth}{@{}p{0.22\linewidth}YY@{}}
    \toprule
    Property & Full pseudo-label mode & Human-label-improvement mode \\
    \midrule
    Human transcript & Not available & Available \\
    ASR path & Language-aware primary and recovery systems & Qwen3-Omni for supported languages; aligned human label otherwise \\
    Quality reference & A second ASR model & The human transcript \\
    Main purpose & Create a transcript & Improve output form while preserving trusted content \\
    \bottomrule
  \end{tabularx}
\end{table}

%% file: figures/text_pipeline.tex
\begin{figure}[b]
\centering
\resizebox{\textwidth}{!}{%
\begin{tikzpicture}[
  font=\small,
  box/.style={draw, rounded corners=3pt, thick, inner sep=5pt,
              text width=#1, align=center, minimum height=1.0cm},
  box/.default=2.4cm,
  decision/.style={draw=orangebdr, diamond, thick, fill=orangebg,
                   aspect=2.0, align=center, inner sep=2pt},
  arw/.style={-{Stealth[length=5pt]}, thick},
  optional/.style={-{Stealth[length=5pt]}, thick, dashed},
]
  \node[box=2.5cm, fill=graybg, draw=graybdr] (input)
    {\textbf{Speech-faithful transcript}};
  \node[box=3.1cm, fill=bluebg, draw=bluebdr, right=0.65cm of input] (core)
    {\textbf{Core text stages}\\[-1pt]{\scriptsize transcript forms, code-switched text, captions, SpeechQA}};
  \node[box=2.6cm, fill=orangebg, draw=orangebdr, right=0.65cm of core] (quality)
    {\textbf{Quality filtering}\\[-1pt]{\scriptsize rules and model agreement}};
  \node[decision, right=0.8cm of quality] (hq) {High\\quality?};
  \node[box=2.6cm, fill=greenbg, draw=greenbdr, right=0.8cm of hq] (release)
    {\textbf{Core release labels}\\[-1pt]{\scriptsize text targets and quality flag}};

  \node[box=2.8cm, fill=purplebg, draw=purplebdr,
        below=1.0cm of quality, xshift=1.1cm] (context)
    {\textbf{Contextual-ASR labels}\\[-1pt]{\scriptsize terms and distractors}};
  \node[box=2.8cm, fill=purplebg, draw=purplebdr, right=0.65cm of context] (translation)
    {\textbf{Cross-lingual translation}\\[-1pt]{\scriptsize spoken content preserved}};
  \node[box=3.0cm, fill=graybg, draw=graybdr, right=0.65cm of translation] (packing)
    {\textbf{Instruction packing}\\[-1pt]{\scriptsize optional training adapter}};

  \draw[arw] (input) -- (core);
  \draw[arw] (core) -- (quality);
  \draw[arw] (quality) -- (hq);
  \draw[arw] (hq) -- node[above, font=\scriptsize] {all decisions} (release);
  \coordinate (gate) at ($(hq.south)+(0,-0.45)$);
  \draw[optional] (hq.south) -- node[right, font=\scriptsize] {yes} (gate) -| (context.north);
  \draw[optional] (gate) -| (translation.north);
  \draw[optional] (context.south) -- ++(0,-0.5) -| (packing.south);
  \draw[optional] (translation) -- (packing);
\end{tikzpicture}%
}
\caption{Text and supervision pipeline. Quality decisions are retained for processed records. Core text stages require a valid transcript. Contextual-ASR labels use selected high-quality records. Most later translation runs use the same selection; early translation runs used the full data. Instruction packing is an optional downstream adapter.}
\label{fig:text-pipeline}
\end{figure}
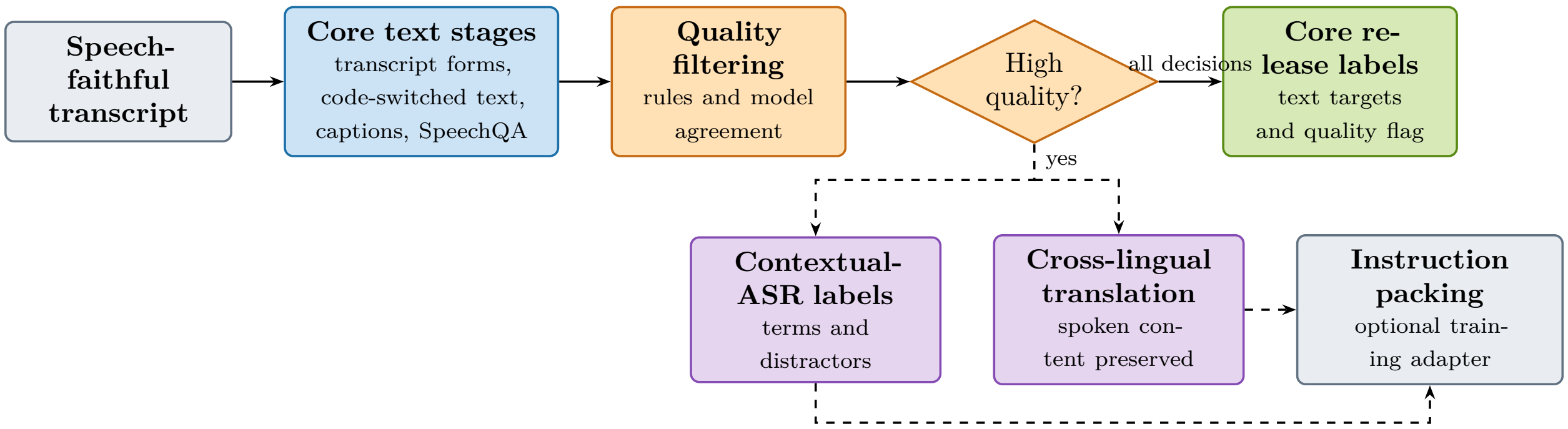

%% file: tables/output_targets.tex
\begin{table}[t]
\centering
  \caption{Curated output targets and their intended behavior.}
  \label{tab:output-targets}
  \centering
  \small
  \begin{tabularx}{\linewidth}{@{}p{0.25\linewidth}YY@{}}
    \toprule
    Output target & Required behavior & Main use \\
    \midrule
    Speech-faithful normalized text & Preserve spoken words, number forms, grammar, and disfluencies & Verbatim ASR and translation source \\
    Written form with disfluencies & Restore written conventions while retaining speech artifacts & Readable, speech-faithful transcripts \\
    Written form without disfluencies & Restore written conventions and remove fillers, repetitions, and false starts & Clean display and downstream text use \\
    Code-switched text & Restore English spelling while retaining native-language words, suffixes, and disfluencies & ASR with an explicit script convention \\
    \bottomrule
  \end{tabularx}
\end{table}

%% file: tables/context_variants.tex
\begin{table}[t]
\centering\small
\caption{Context variants in the reference training adapter. The transcript target stays tied to the audio.}
\label{tab:context-variants}
\begin{tabularx}{\linewidth}{@{}p{0.18\linewidth}Y@{}}
\toprule
Variant & Information supplied in the prompt \\
\midrule
Coarse & Domain label \\
Fine & Domain and relevant terms \\
Distractor & Relevant terms mixed with irrelevant terms \\
Partial & A subset of relevant terms \\
Negative & An unrelated domain, with the reference variant's term selection \\
\bottomrule
\end{tabularx}
\end{table}

%% file: tables/instruction_targets.tex
\begin{table}[H]
\centering\small
\caption{Instruction families and their targets in the reference packer. A pair requires a usable target.}
\label{tab:instruction-targets}
\begin{tabularx}{\linewidth}{@{}p{0.30\linewidth}Y@{}}
\toprule
Instruction family & Target \\
\midrule
Transcription & Prefer \path|tn_raw|; use the configured transcript fallback when absent \\
Written form & \path|itn_raw| \\
Written form without disfluencies & \path|itn_clean| \\
Code-switched ASR & \path|code_switched_text| with the requested script convention \\
Captioning & \path|captioning_text| \\
SpeechQA & Parsed question as instruction; parsed answer as target \\
Contextual ASR & Context prompt paired with the transcript target \\
Speech translation & Target-language \path|translation_raw| passing the configured score threshold \\
\bottomrule
\end{tabularx}
\end{table}

%% file: figures/output_example.tex
\begin{figure}[t]
\centering
\includegraphics[width=\linewidth]{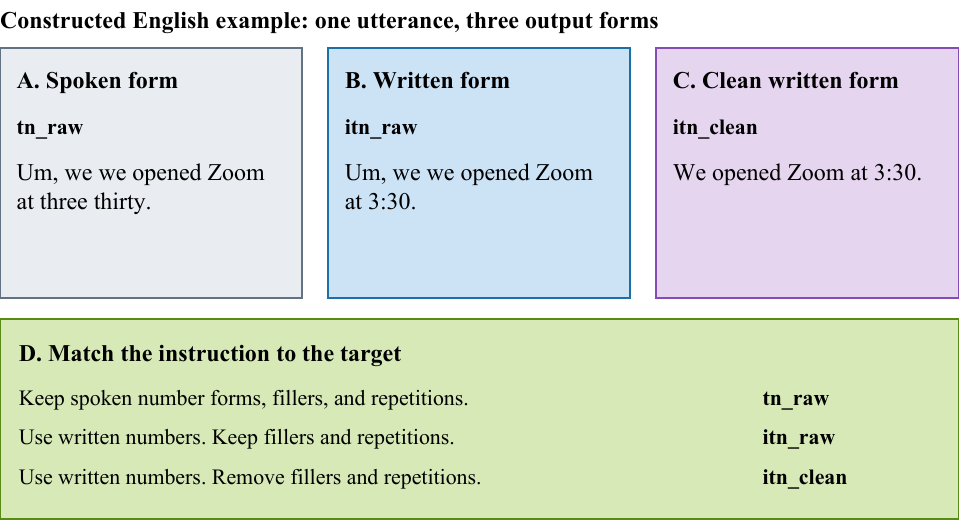}
\par\medskip
\setlength{\fboxsep}{7pt}
\fcolorbox{orangebdr}{orangebg}{%
\parbox[t]{0.935\linewidth}{%
\textbf{E. Code-switched text: constructed Russian example}\par\smallskip
\includegraphics[width=\linewidth]{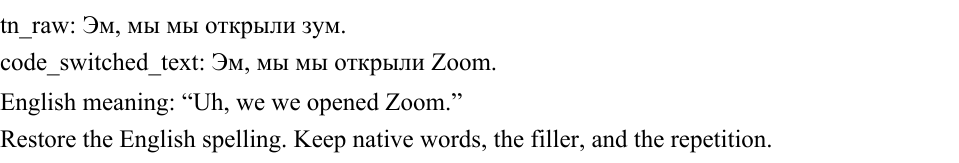}}}
\caption{Constructed examples of output targets and instruction-based selection. Panels A--C show spoken numbers, written numbers, and disfluency removal for the same English utterance. Panel D pairs instructions with the corresponding target fields. Panel E illustrates script restoration in Russian. These examples are illustrative and are not released records.}
\label{fig:output-example}
\end{figure}

%% file: tables/throughput.tex
\begin{table}[H]
\centering\small
\caption{Processing rates for the listed workloads. GPU rates use A100 80~GB allocations.}
\label{tab:throughput}
\begin{tabularx}{\linewidth}{@{}p{0.26\linewidth}p{0.30\linewidth}Y@{}}
\toprule
Stage & Rate & Workload \\
\midrule
Metadata extraction & $\sim$23 audio-h/GPU-h & Depends on recording length and work balance \\
ASR labeling & $\sim$281 audio-h/node-h & Measured short-utterance workload \\
Text enrichment & $\sim$15--25 audio-h/node-h & Range across job sizes; startup affects small jobs \\
Quality filtering & $\sim$10,400 audio-h/CPU-node-h & CPU-only workload \\
Contextual ASR & $\sim$30 audio-h/node-h & Includes the gate, prompt variants, and I/O \\
Translation & $\sim$144 eligible audio-h/node-h & Japanese-to-English only \\
\bottomrule
\end{tabularx}
\end{table}

%% file: sections/pipeline_application.tex
\section{Applying the Pipelines to Granary-v2}
\label{sec:application}

\input{tables/dataset_paths}

The current inputs contain short audio segments. We apply ASR labeling and text processing without a metadata-extraction pass. Table~\ref{tab:dataset-paths} shows the processing paths.

\subsection{Audio-Only ASR Labeling}

We generate new ASR labels for LibriLight, VoxPopuli with MOSEL labels, the government-origin subset of People's Speech, YODAS2, and YouTube-Commons. The primary and recovery models provide candidate transcripts from the audio for agreement checks. These sources describe the processing scope. Labels and instruction pairs from YODAS2, YouTube-Commons, and ReazonSpeech-v2 are excluded from the public release.

\subsection{Corpora with Existing Transcripts}

For inputs with a trusted source transcript, we retain that transcript as a reference. Supported languages use audio-guided Qwen3-Omni label improvement. Other inputs retain the aligned source label. Entity recovery uses the reference to restore trusted name spellings where applicable.

Both paths use the same text stages and quality rules. Context labels are generated for selected high-quality records. Translation selection changed between runs: early runs used the full data, while later runs mainly used high-quality records, as described in Section~\ref{sec:translation}.

%% file: tables/dataset_paths.tex
\begin{table}[b]
\centering
  \caption{Dataset-specific paths through the reusable pipelines.}
  \label{tab:dataset-paths}
  \centering
  \scriptsize
  \begin{tabularx}{\linewidth}{@{}p{0.20\linewidth}p{0.13\linewidth}p{0.20\linewidth}p{0.20\linewidth}Y@{}}
    \toprule
    Input & Metadata & ASR & Text pipeline & Optional enrichment \\
    \midrule
    Audio without a trusted transcript & Skipped & Full pseudo-label mode & Core stages and quality filtering & Context and translation on selected records \\
    Audio with a source transcript & Skipped & Improve or retain the source label & Entity recovery when applicable, core stages, quality filtering & Context and translation on selected records \\
    \bottomrule
  \end{tabularx}
\end{table}

%% file: sections/dataset.tex
\section{The Granary-v2 Dataset}
\label{sec:dataset}

\input{tables/release_summary}

Table~\ref{tab:release-summary} distinguishes the full processed collection from the public release subset~\citep{granary-v2-data}. Appendix~\ref{app:sources} lists the corpus entries, their durations, and their release status.

\paragraph{Public release scope.}
The full processed collection contains 19 source entries. The public release excludes all generated labels and instruction--target pairs derived from YODAS2, YouTube-Commons, and ReazonSpeech-v2. This exclusion covers transcripts, translations, captions, code-switched text, question--answer pairs, and context labels. The remaining 16 entries correspond to 463,062.37 hours of source audio, with 92,495.56 hours of translation into English. These subset totals are calculated from the corpus durations reported here. The Hugging Face repository distributes label manifests. Source audio is obtained separately from its distributors.

\paragraph{Counting conventions.}
The full processed collection contains 843,579.22 hours of unique ASR audio, of which 317,724 hours carry the high-quality flag. This total is the sum of unique audio durations across the source corpora. The two VoxPopuli variants remain separate entries, and LibriSpeech-PC is the sole LibriSpeech contribution. Transcript variants and translations are labels attached to these recordings and do not increase the audio-hour total. Translation runs in both directions: 142,778.43 hours of non-English audio are translated into English, and 55,772 hours of English audio are translated into 31 target languages at once. Both directions count distinct audio once per record, however many targets that record carries, so neither double-counts and the two never sum past a corpus's hours. Translation coverage includes early runs on the full data and later runs that mainly used high-quality records. We use only the training splits of the source corpora. Source validation and test splits are excluded from the training data.

\subsection{Language, Source, and Duration Analysis}
\label{sec:lang-source-duration}

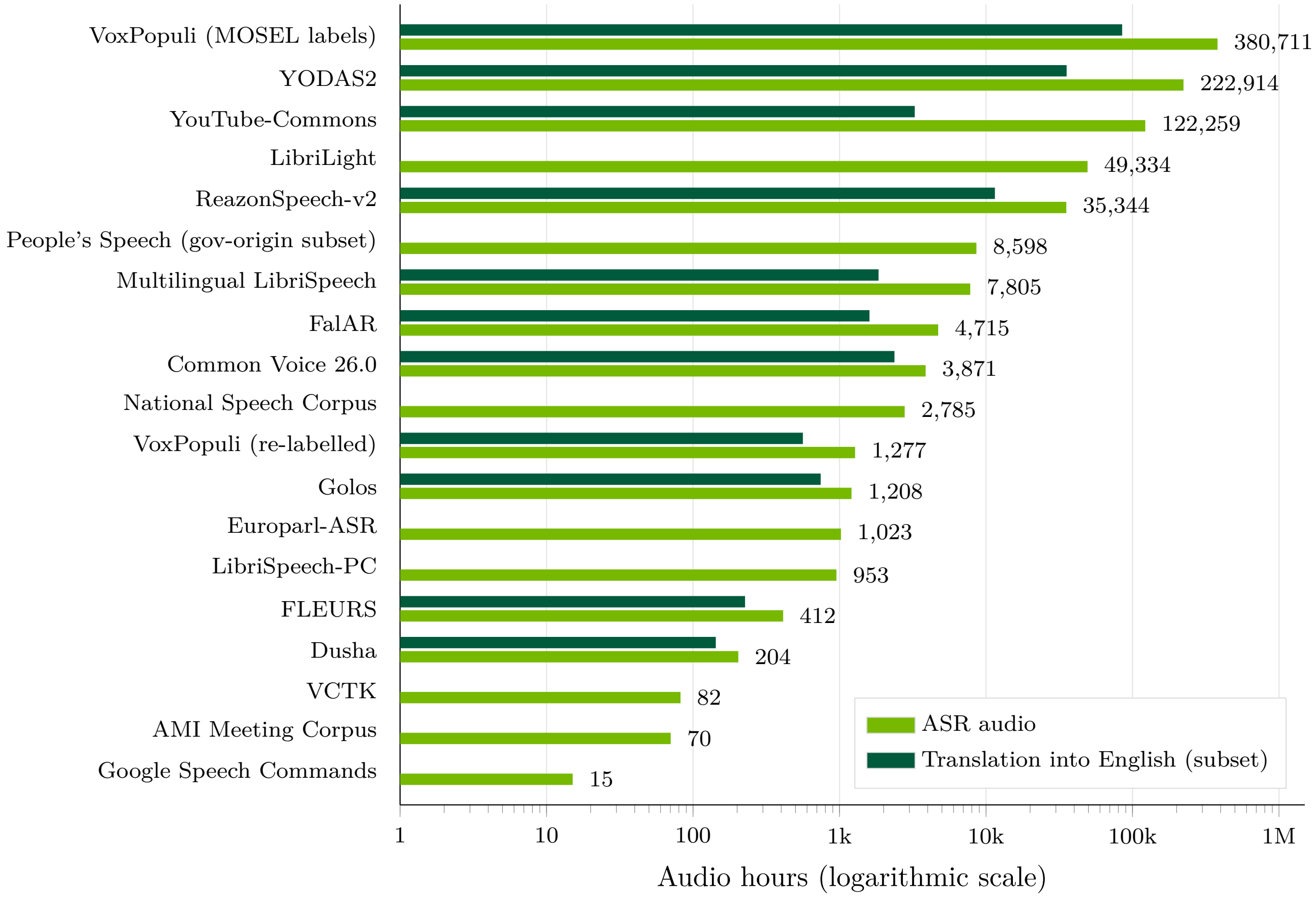
\begin{figure}[p]
\centering
\input{figures/corpus-hours}
\caption{Unique ASR audio hours and translation coverage by corpus entry in the full processed collection~\citep{granary-v2-data}. YODAS2, YouTube-Commons, and ReazonSpeech-v2 are included in these statistics but excluded from the public release. Translation into English uses a subset of the ASR audio. Bar labels show ASR hours rounded to the nearest hour.}
\label{fig:corpus-hours}
\end{figure}

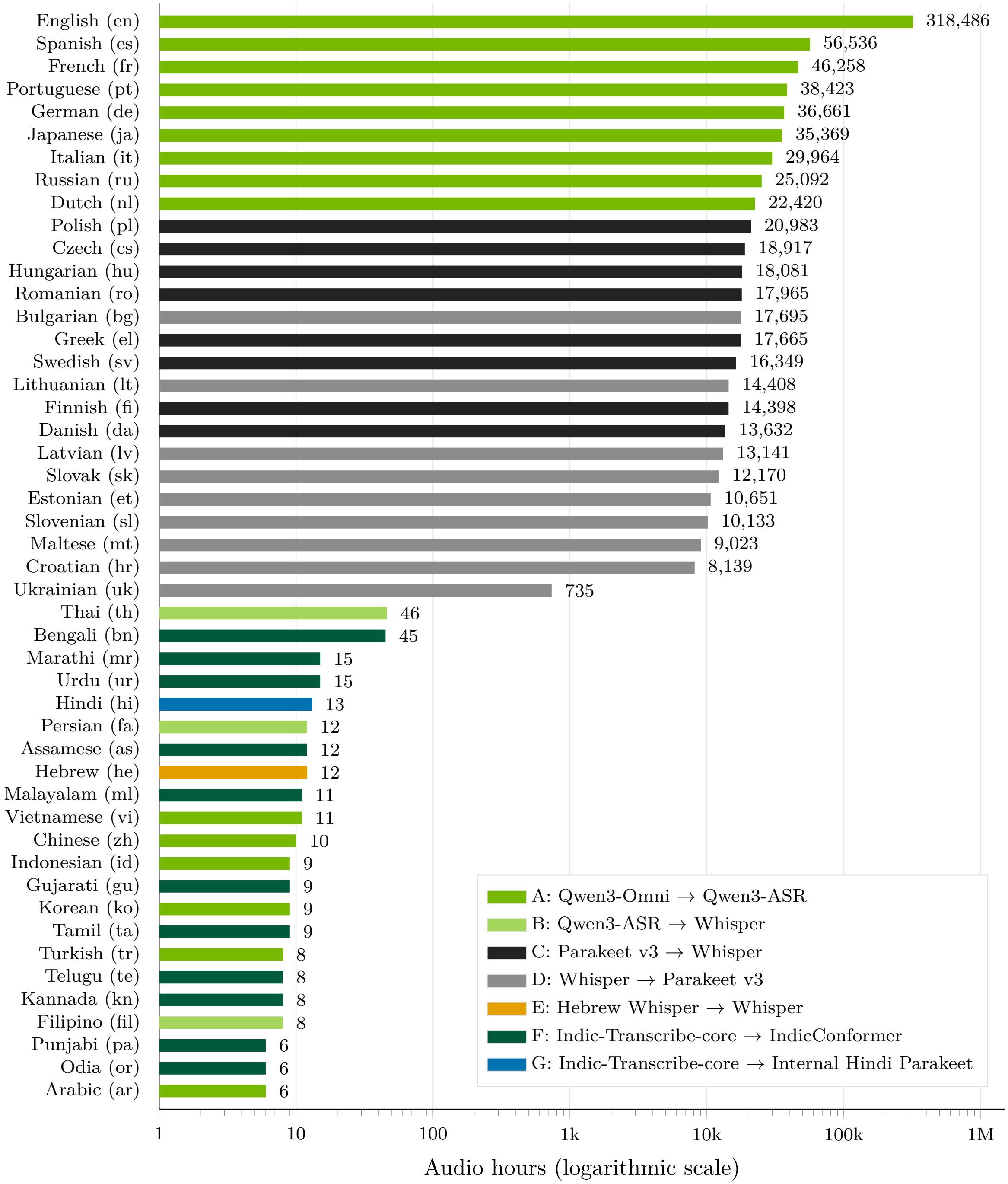
\begin{figure}[p]
\centering
\input{figures/language-hours}
\caption{ASR hours by language in the full processed collection, using the previously reported rounded values~\citep{granary-v2-data}. These statistics include YODAS2, YouTube-Commons, and ReazonSpeech-v2, which are excluded from the public release. Colors identify the primary and recovery model routes in Table~\ref{tab:asr-groups}. English includes only LibriSpeech-PC for the LibriSpeech contribution.}
\label{fig:language-hours}
\end{figure}

Figures~\ref{fig:corpus-hours} and~\ref{fig:language-hours} show the corpus and language distributions of the full processed collection. The language chart also shows the ASR model routes. VoxPopuli with MOSEL labels, YODAS2, YouTube-Commons, LibriLight, and ReazonSpeech-v2 account for about 96.1\% of reported ASR hours. The remaining 14 entries account for about 3.9\%. This concentration makes corpus and language sampling important during training.

\subsection{Label and Task Coverage}

Labels include transcript variants, captions, code-switched text, and question--answer pairs. Selected records also receive context and translations. Available targets depend on stage eligibility and validation. Instruction--target pairs provide a training view of these labels.

%% file: tables/release_summary.tex
\begin{table}[b]
\centering\small
\caption{Full processed collection and public release subset. Translation hours refer to audio already counted for ASR. Public subset hours are calculated by excluding YODAS2, YouTube-Commons, and ReazonSpeech-v2 from the reported corpus durations. A dash marks a subset statistic not separately counted.}
\label{tab:release-summary}
\setlength{\tabcolsep}{4pt}
\begin{tabular}{@{}lrr@{}}
\toprule
Measure & Full processed & Public subset \\
\midrule
Corpus entries & 19 & 16 \\
Languages & 48 & 48 \\
Unique ASR hours & 843,579.22 & 463,062.37 \\
Utterance records & $\sim$213.4 million & -- \\
High-quality ASR hours & 317,724 & -- \\
Translation into English, hours & 142,778.43 & 92,495.56 \\
Translation out of English, hours & 55,772 & -- \\
\bottomrule
\end{tabular}
\end{table}

%% file: figures/corpus-hours.tex
\begin{tikzpicture}
\definecolor{asrhours}{HTML}{76B900}
\definecolor{translationhours}{HTML}{005A3C}
\begin{axis}[
width=0.775\linewidth, height=11.5cm,
xbar, xmode=log, xmin=1, xmax=1500000,
ymin=-0.8, ymax=18.8,
ytick={0,1,2,3,4,5,6,7,8,9,10,11,12,13,14,15,16,17,18}, yticklabels={{Google Speech Commands},{AMI Meeting Corpus},{VCTK},{Dusha},{FLEURS},{LibriSpeech-PC},{Europarl-ASR},{Golos},{VoxPopuli (re-labelled)},{National Speech Corpus},{Common Voice 26.0},{FalAR},{Multilingual LibriSpeech},{People's Speech (gov-origin subset)},{ReazonSpeech-v2},{LibriLight},{YouTube-Commons},{YODAS2},{VoxPopuli (MOSEL labels)}},
yticklabel style={font=\scriptsize},
xtick={1,10,100,1000,10000,100000,1000000},
xticklabels={1,10,100,1k,10k,100k,1M},
xticklabel style={font=\scriptsize},
xlabel={Audio hours (logarithmic scale)}, xlabel style={font=\small},
axis x line*=bottom, axis y line*=left,
tick align=outside, ytick style={draw=none},
xmajorgrids=true, grid style={gray!20}, bar width=4pt,
legend style={at={(0.98,0.02)}, anchor=south east,
font=\scriptsize, draw=gray!25, fill=white, inner sep=4pt,
legend cell align=left, row sep=1pt},
]
\addplot[area legend,fill=asrhours,draw=none,bar shift=-2.5pt,
point meta=explicit,
nodes near coords={\pgfmathprintnumber[fixed,precision=0,1000 sep={\mathord{,}}]{\pgfplotspointmeta}},
every node near coord/.append style={font=\scriptsize,anchor=west,xshift=2pt},
] coordinates {
(15.08,0)[15.08]
(70.38,1)[70.38]
(82.15,2)[82.15]
(203.73,3)[203.73]
(412.18,4)[412.18]
(952.88,5)[952.88]
(1022.70,6)[1022.70]
(1207.72,7)[1207.72]
(1276.96,8)[1276.96]
(2784.60,9)[2784.60]
(3871.23,10)[3871.23]
(4714.83,11)[4714.83]
(7804.56,12)[7804.56]
(8597.76,13)[8597.76]
(35344.08,14)[35344.08]
(49334.35,15)[49334.35]
(122258.88,16)[122258.88]
(222913.89,17)[222913.89]
(380711.26,18)[380711.26]
};
\addlegendentry{ASR audio}
\addplot[area legend,fill=translationhours,draw=none,bar shift=2.5pt] coordinates {
(143.01,3)
(226.48,4)
(743.57,7)
(562.21,8)
(2371.89,10)
(1601.57,11)
(1847.54,12)
(11503.98,14)
(3261.71,16)
(35517.18,17)
(84999.29,18)
};
\addlegendentry{Translation into English (subset)}
\end{axis}
\end{tikzpicture}

%% file: figures/language-hours.tex
\begin{tikzpicture}
\definecolor{routeA}{HTML}{76B900}
\definecolor{routeB}{HTML}{A5D65C}
\definecolor{routeC}{HTML}{232323}
\definecolor{routeD}{HTML}{8C8C8C}
\definecolor{routeE}{HTML}{E69F00}
\definecolor{routeF}{HTML}{005A3C}
\definecolor{routeG}{HTML}{0072B2}
\begin{axis}[
width=0.88\linewidth, height=18.5cm,
xbar, xmode=log, xmin=1, xmax=1500000,
ymin=-0.8, ymax=47.8,
ytick={0,1,2,3,4,5,6,7,8,9,10,11,12,13,14,15,16,17,18,19,20,21,22,23,24,25,26,27,28,29,30,31,32,33,34,35,36,37,38,39,40,41,42,43,44,45,46,47}, yticklabels={{Arabic (ar)},{Odia (or)},{Punjabi (pa)},{Filipino (fil)},{Kannada (kn)},{Telugu (te)},{Turkish (tr)},{Tamil (ta)},{Korean (ko)},{Gujarati (gu)},{Indonesian (id)},{Chinese (zh)},{Vietnamese (vi)},{Malayalam (ml)},{Hebrew (he)},{Assamese (as)},{Persian (fa)},{Hindi (hi)},{Urdu (ur)},{Marathi (mr)},{Bengali (bn)},{Thai (th)},{Ukrainian (uk)},{Croatian (hr)},{Maltese (mt)},{Slovenian (sl)},{Estonian (et)},{Slovak (sk)},{Latvian (lv)},{Danish (da)},{Finnish (fi)},{Lithuanian (lt)},{Swedish (sv)},{Greek (el)},{Bulgarian (bg)},{Romanian (ro)},{Hungarian (hu)},{Czech (cs)},{Polish (pl)},{Dutch (nl)},{Russian (ru)},{Italian (it)},{Japanese (ja)},{German (de)},{Portuguese (pt)},{French (fr)},{Spanish (es)},{English (en)}},
yticklabel style={font=\scriptsize},
xtick={1,10,100,1000,10000,100000,1000000},
xticklabels={1,10,100,1k,10k,100k,1M},
xticklabel style={font=\scriptsize},
xlabel={Audio hours (logarithmic scale)}, xlabel style={font=\small},
axis x line*=bottom, axis y line*=left,
tick align=outside, ytick style={draw=none},
xmajorgrids=true, grid style={gray!20}, bar width=5.5pt,
legend style={at={(0.98,0.02)}, anchor=south east,
font=\scriptsize, draw=gray!25, fill=white, inner sep=4pt,
legend cell align=left, row sep=1pt},
]
\addplot[area legend,fill=routeA,draw=none,bar shift=0pt,
point meta=explicit,
nodes near coords={\pgfmathprintnumber[fixed,precision=0,1000 sep={\mathord{,}}]{\pgfplotspointmeta}},
every node near coord/.append style={font=\scriptsize,anchor=west,xshift=2pt},
] coordinates {
(6,0)[6]
(8,6)[8]
(9,8)[9]
(9,10)[9]
(10,11)[10]
(11,12)[11]
(22420,39)[22420]
(25092,40)[25092]
(29964,41)[29964]
(35369,42)[35369]
(36661,43)[36661]
(38423,44)[38423]
(46258,45)[46258]
(56536,46)[56536]
(318486,47)[318486]
};
\addlegendentry{A: Qwen3-Omni $\rightarrow$ Qwen3-ASR}
\addplot[area legend,fill=routeB,draw=none,bar shift=0pt,
point meta=explicit,
nodes near coords={\pgfmathprintnumber[fixed,precision=0,1000 sep={\mathord{,}}]{\pgfplotspointmeta}},
every node near coord/.append style={font=\scriptsize,anchor=west,xshift=2pt},
] coordinates {
(8,3)[8]
(12,16)[12]
(46,21)[46]
};
\addlegendentry{B: Qwen3-ASR $\rightarrow$ Whisper}
\addplot[area legend,fill=routeC,draw=none,bar shift=0pt,
point meta=explicit,
nodes near coords={\pgfmathprintnumber[fixed,precision=0,1000 sep={\mathord{,}}]{\pgfplotspointmeta}},
every node near coord/.append style={font=\scriptsize,anchor=west,xshift=2pt},
] coordinates {
(13632,29)[13632]
(14398,30)[14398]
(16349,32)[16349]
(17665,33)[17665]
(17965,35)[17965]
(18081,36)[18081]
(18917,37)[18917]
(20983,38)[20983]
};
\addlegendentry{C: Parakeet v3 $\rightarrow$ Whisper}
\addplot[area legend,fill=routeD,draw=none,bar shift=0pt,
point meta=explicit,
nodes near coords={\pgfmathprintnumber[fixed,precision=0,1000 sep={\mathord{,}}]{\pgfplotspointmeta}},
every node near coord/.append style={font=\scriptsize,anchor=west,xshift=2pt},
] coordinates {
(735,22)[735]
(8139,23)[8139]
(9023,24)[9023]
(10133,25)[10133]
(10651,26)[10651]
(12170,27)[12170]
(13141,28)[13141]
(14408,31)[14408]
(17695,34)[17695]
};
\addlegendentry{D: Whisper $\rightarrow$ Parakeet v3}
\addplot[area legend,fill=routeE,draw=none,bar shift=0pt,
point meta=explicit,
nodes near coords={\pgfmathprintnumber[fixed,precision=0,1000 sep={\mathord{,}}]{\pgfplotspointmeta}},
every node near coord/.append style={font=\scriptsize,anchor=west,xshift=2pt},
] coordinates {
(12,14)[12]
};
\addlegendentry{E: Hebrew Whisper $\rightarrow$ Whisper}
\addplot[area legend,fill=routeF,draw=none,bar shift=0pt,
point meta=explicit,
nodes near coords={\pgfmathprintnumber[fixed,precision=0,1000 sep={\mathord{,}}]{\pgfplotspointmeta}},
every node near coord/.append style={font=\scriptsize,anchor=west,xshift=2pt},
] coordinates {
(6,1)[6]
(6,2)[6]
(8,4)[8]
(8,5)[8]
(9,7)[9]
(9,9)[9]
(11,13)[11]
(12,15)[12]
(15,18)[15]
(15,19)[15]
(45,20)[45]
};
\addlegendentry{F: Indic-Transcribe-core $\rightarrow$ IndicConformer}
\addplot[area legend,fill=routeG,draw=none,bar shift=0pt,
point meta=explicit,
nodes near coords={\pgfmathprintnumber[fixed,precision=0,1000 sep={\mathord{,}}]{\pgfplotspointmeta}},
every node near coord/.append style={font=\scriptsize,anchor=west,xshift=2pt},
] coordinates {
(13,17)[13]
};
\addlegendentry{G: Indic-Transcribe-core $\rightarrow$ Internal Hindi Parakeet}
\end{axis}
\end{tikzpicture}

%% file: sections/evaluation.tex
\section{Model Training and Downstream Results}
\label{sec:evaluation}

Granary-v2 supplies ASR training labels and instruction--target pairs for speech-language models. We compare English ASR training with all labels and with the high-quality subset. We then summarize downstream ASR and translation results from Nemotron 3.5 Transcribe. The reported training mixtures and results describe the experiments as run. They are separate from the public release subset defined in Section~\ref{sec:dataset}. The release exclusion does not change these training statistics or scores.

\subsection{Transformer-TDT Training}

\begin{table}[b]
\centering
\small
\caption{English Transformer-TDT results with high-quality and all labels. Scores are WER (\%); lower is better. Average WER is the unweighted mean across the nine benchmarks. Bold indicates the better score for each benchmark.}
\label{tab:english-quality-results}
\begin{tabular}{@{}lrr@{}}
\toprule
Benchmark & High-quality labels & All labels \\
\midrule
AMI-Cleaned~\citep{carletta2005ami} & \textbf{7.84} & 8.72 \\
Earnings22-Cleaned-AA-chunked~\citep{delrio2022earnings22} & \textbf{5.77} & 6.15 \\
Gigaspeech-Cleaned~\citep{chen2021gigaspeech} & \textbf{7.33} & 7.41 \\
LS Clean~\citep{panayotov2015librispeech} & 1.13 & \textbf{1.10} \\
LS Other~\citep{panayotov2015librispeech} & 2.67 & \textbf{2.61} \\
SPGISpeech~\citep{oneill2021spgispeech} & \textbf{2.74} & 3.08 \\
Voice Arena Monsoon~\citep{voicearena2026monsoon} & \textbf{3.33} & 3.41 \\
Voxpopuli-AA-Cleaned~\citep{voxpopuli} & 1.68 & \textbf{1.57} \\
URGENT24~\citep{zhang2024urgent} & \textbf{9.56} & 10.11 \\
\midrule
Average WER                 & \textbf{4.67} & 4.91 \\
\bottomrule
\end{tabular}
\end{table}

\label{sec:tdt-training}

To evaluate the effectiveness of quality filtering, we train two English
Transformer-TDT models with approximately 600M-parameter encoders: one
using all training data and the other using only the subset marked as
high quality. Both models use the same architecture and training recipe
and are trained for 300k steps, allowing us to compare the two data
selections under a fixed training-step budget. We use the Token-and-Duration Transducer (TDT) objective~\citep{tdt}.

The English training mixture contains 347,178.90 audio hours before
quality filtering. The high-quality subset contains 162,845.95 hours,
or 46.91\% of the mixture. These durations follow the flat English
training configuration. This mixture includes additional corpora beyond
the full processed Granary-v2 collection, which contains approximately
318,486 English audio hours before filtering in
Figure~\ref{fig:language-hours}. This figure includes sources excluded
from the public release; it is not the public subset's English duration.

Training uses BF16 precision and AdamW with a peak learning rate of
$5 \times 10^{-4}$, betas $(0.9, 0.95)$, and weight decay of $10^{-2}$.
A cosine learning-rate schedule uses 15,000 warmup steps and decays to
a minimum learning rate of $10^{-6}$. Both runs use
\texttt{tn\_raw} as the training target. Detailed architecture,
acoustic feature, and tokenizer settings are provided in
Appendix~\ref{app:training-recipes}.

Table~\ref{tab:english-quality-results} reports WER on the English ASR
leaderboard benchmarks~\citep{srivastav2025openasr}. The high-quality subset yields lower WER on
six of the nine reported benchmarks, with the largest improvements
on AMI, Earnings22, and SPGISpeech. Training on all data performs
slightly better on LibriSpeech Clean, LibriSpeech Other, and VoxPopuli.
Across these nine benchmarks, the unweighted mean WER decreases
from 4.91\% to 4.67\%. These results show that quality filtering
improves average recognition accuracy under the same training-step
budget, despite retaining less training data.

\subsection{Downstream Use in Nemotron 3.5 Transcribe}
\label{sec:downstream-use}

\input{tables/downstream_results}

Nemotron 3.5 Transcribe initializes its ASR encoder from a multilingual TDT pretrained on Granary-v2~\citep{nemotron35-transcribe}. The encoder has about 600M parameters and produces one representation per 80~ms. Transcribe jointly trains this encoder, a projection, and a language-model decoder on a mixture of speech datasets, followed by reinforcement learning. Granary-v2 supplies transcript, context, question--answer, and translation targets for multi-task training.

Table~\ref{tab:downstream-results} summarizes the ASR and translation results from the Transcribe technical report. The seven-language leaderboard comparison uses a baseline snapshot dated 2 October 2026. The standalone FLEURS evaluation~\citep{conneau2022fleurs} uses a separate protocol. Translation is scored against the benchmark references with BLEU~\citep{bleu} or spBLEU~\citep{flores101}. The source report uses SacreBLEU with language-specific tokenizers~\citep{sacrebleu}. Thai targets use the FLORES-200 SentencePiece tokenizer~\citep{nllb}.

Appendix~\ref{app:downstream-results} gives per-language scores and evaluation instructions.

%% file: tables/downstream_results.tex
\begin{table}[H]
\centering\small
\caption{ASR and speech translation results for \transcribename{}, trained on a mixture that includes Granary-v2~\citep{nemotron35-transcribe}.}
\label{tab:downstream-results}
\begin{tabularx}{\linewidth}{@{}Ylrl@{}}
\toprule
Evaluation & Languages & Score & Metric \\
\midrule
Multilingual ASR leaderboard & 7 & 4.62 & WER (\%)$^\dagger$ \\
FLEURS ASR~\citep{conneau2022fleurs} & 37 & 9.29 & Mean WER/CER (\%) \\
CoVoST~2 ASR~\citep{covost2} & 17 & 12.19 & Mean WER/CER (\%) \\
FLEURS: other languages to English & 36 & 29.20 & Mean BLEU \\
FLEURS: English to other languages & 36 & 26.54 & Mean BLEU$^\ast$ \\
CoVoST~2: other languages to English & 16 & 33.67 & Mean BLEU \\
CoVoST~2: English to other languages & 10 & 26.34 & Mean BLEU \\
\bottomrule
\end{tabularx}
\par\smallskip
\noindent\begin{minipage}{\linewidth}\footnotesize
Lower error rates and higher translation scores are better.
$^\dagger$Hindi uses orthographically informed WER on the public test set.
$^\ast$Thai targets use spBLEU. Other translation targets retain the source report's tokenizers.
The leaderboard and standalone FLEURS evaluations use different protocols.
The 37-language FLEURS result includes Norwegian, which the source report does not list as officially supported for transcription.
\end{minipage}
\end{table}

%% file: sections/responsible_release.tex
\section{Limitations and Release Conditions}
\label{sec:limitations}

Language and source coverage are uneven. Teacher models can share transcription errors, and source transcripts can contain recognition or formatting errors. Entity recovery, normalization, context generation, and translation can introduce unsupported changes. Text-only stages infer these changes from transcripts. Audio-based assessment is needed for ambiguous spoken forms, disfluencies, and code-switched words. Generated captions can use a different language from the requested target language.

The English filtering experiment covers one ASR architecture and a fixed training-step budget. It measures label filtering and does not compare individual instruction families. The instruction--target pairs provide supervision for prompt-based output selection in speech-language models. The text-model comparison uses qualitative samples from three languages. Processing rates depend on workload, hardware, and timing boundaries. Reference prompts and launch settings describe the configurations supplied with the report.

The public release excludes all labels and instruction--target pairs derived from YODAS2, YouTube-Commons, and ReazonSpeech-v2. Their statistics describe processing and experimental scope only; they do not indicate public availability. The repository distributes manifests rather than source audio.

The release uses source training splits and excludes source validation and test splits from training. Audio, generated labels, model checkpoints, and software have their own access and license terms. The dataset repository is \href{https://huggingface.co/datasets/nvidia/Granary-v2}{\texttt{nvidia/Granary-v2}}~\citep{granary-v2-data}. The high-quality flag measures structural validity and transcript agreement; it is not a privacy or harmful-content label.

%% file: sections/conclusion.tex
\section{Conclusion}
\label{sec:conclusion}

Granary-v2 processes 843,579.22 unique ASR hours across 48 languages. The public subset excludes labels and instruction pairs derived from YODAS2, YouTube-Commons, and ReazonSpeech-v2. It contains labels for 16 source entries, corresponding to 463,062.37 hours of source audio. Aligned transcript forms and instructions support prompt-based output selection. At 300,000 training steps, high-quality labels reduce English Transformer-TDT mean WER from 4.91\% to 4.67\% across nine benchmarks. The data also supports encoder pretraining and multi-task training in Nemotron 3.5 Transcribe.

%% file: sections/contributors.tex
\section*{Contributors}
\addcontentsline{toc}{section}{Contributors}
\label{sec:contributors}

Nithin Rao Koluguri\textsuperscript{*}, Meline Mkrtchyan\textsuperscript{*}, Nune Tadevosyan\textsuperscript{*}, Aaftab V, Sushmitha Deva, Fan Qian, Davit Karamyan, Kunal Dhawan, Taejin Park, Piotr {\.Z}elasko, Arushi Dhingra, Viraj Karandikar, Shuoyang Ding, Slyne Deng, Onur Yilmaz, Praateek Mahajan, Utkarsh Vaidya, Nikolay Karpov, Jagadeesh Balam, Boris Ginsburg

{\small \textit{Authors marked * contributed equally and are listed alphabetically.}\par}

%% file: sections/appendices.tex
\section{Corpus Entries and Translation Coverage}
\label{app:sources}

\input{tables/source_corpora}

Table~\ref{tab:source-corpora} lists corpus groups, language counts, and durations for the full processed collection~\citep{granary-v2-data}. The public-release column identifies which entries have released labels. YODAS2, YouTube-Commons, and ReazonSpeech-v2 remain in the processing statistics but have no labels or instruction pairs in the public release. VoxPopuli with MOSEL labels and re-labelled VoxPopuli remain separate entries. LibriSpeech-PC is the sole LibriSpeech contribution. Translation hours refer to labels attached to the ASR recordings.

\section{Pipeline Configuration and Recovery}
\label{app:pipeline-config}

Metadata extraction runs diarization on the full recording before VAD segmentation. Segment speaker counts are derived from the retained RTTM timeline. The output manifest records successful segments, empty VAD results, and read errors. Completion checks account for every unique input recording, including empty VAD results and read errors.

ASR and text stages retain intermediate outputs in separate fields. A job can resume at a completed shard or stage. Row-count checks validate completion markers. GPU placement and batching depend on the model. CPU filtering and instruction packing can run separately from model inference.

For People's Speech, a 3,032-shard run reached 61\% completion in its first time window. Dependent resume jobs reached 100\% without repeating completed work.

\input{sections/appendix_reference_config}

\section{Multilingual ASR Routing}
\label{sec:multilingual-routing}

\begin{table}[H]
  \centering
  \small
  \caption{Configured language membership of the ASR routing groups in Table~\ref{tab:asr-groups}.}
  \label{tab:asr-group-membership}
  \begin{tabular}{@{}crp{0.72\linewidth}@{}}
    \toprule
    Group & \# & Languages \\
    \midrule
    \textbf{A} & 15 & Arabic~(\texttt{ar}), Chinese~(\texttt{zh}), Dutch~(\texttt{nl}), English~(\texttt{en}), French~(\texttt{fr}), German~(\texttt{de}), Indonesian~(\texttt{id}), Italian~(\texttt{it}), Japanese~(\texttt{ja}), Korean~(\texttt{ko}), Portuguese~(\texttt{pt}), Russian~(\texttt{ru}), Spanish~(\texttt{es}), Turkish~(\texttt{tr}), Vietnamese~(\texttt{vi}) \\
    \addlinespace
    \textbf{B} & 3 & Filipino~(\texttt{fil}), Persian~(\texttt{fa}), Thai~(\texttt{th}) \\
    \addlinespace
    \textbf{C} & 8 & Czech~(\texttt{cs}), Danish~(\texttt{da}), Finnish~(\texttt{fi}), Greek~(\texttt{el}), Hungarian~(\texttt{hu}), Polish~(\texttt{pl}), Romanian~(\texttt{ro}), Swedish~(\texttt{sv}) \\
    \addlinespace
    \textbf{D} & 9 & Bulgarian~(\texttt{bg}), Croatian~(\texttt{hr}), Estonian~(\texttt{et}), Latvian~(\texttt{lv}), Lithuanian~(\texttt{lt}), Maltese~(\texttt{mt}), Slovak~(\texttt{sk}), Slovenian~(\texttt{sl}), Ukrainian~(\texttt{uk}) \\
    \addlinespace
    \textbf{E} & 1 & Hebrew~(\texttt{he}) \\
    \addlinespace
    \textbf{F} & 11 & Assamese~(\texttt{as}), Bengali~(\texttt{bn}), Gujarati~(\texttt{gu}), Kannada~(\texttt{kn}), Malayalam~(\texttt{ml}), Marathi~(\texttt{mr}), Odia~(\texttt{or}), Punjabi~(\texttt{pa}), Tamil~(\texttt{ta}), Telugu~(\texttt{te}), Urdu~(\texttt{ur}) \\
    \addlinespace
    \textbf{G} & 1 & Hindi~(\texttt{hi}) \\
    \bottomrule
  \end{tabular}
\end{table}

\begin{table}[H]
\centering
\small
\caption{Checkpoint identifiers for the ASR routing groups. Internal Hindi Parakeet is an internal checkpoint.}
\label{tab:asr-model-aliases}
\begin{tabularx}{\linewidth}{@{}p{0.26\linewidth}Y@{}}
\toprule
Model & Checkpoint or implementation \\
\midrule
Qwen3-Omni~\citep{xu2025qwen3omni} & \path|Qwen/Qwen3-Omni-30B-A3B-Instruct| \\
Qwen3-ASR~\citep{shi2026qwen3asr} & \path|Qwen/Qwen3-ASR-0.6B| \\
Parakeet v3~\citep{sekoyan2025canaryv2,parakeet-v3} & \path|nvidia/parakeet-tdt-0.6b-v3| \\
Whisper~\citep{radford2022whisper,whisper-large-v3} & Whisper Large v3 through faster-whisper~\citep{systran_fasterwhisper} \\
Hebrew Whisper~\citep{hebrew-whisper} & \path|ivrit-ai/whisper-large-v3| \\
Indic-Transcribe-core~\citep{indic-transcribe-core} & \href{https://huggingface.co/bodhan-ai/indic-transcribe-core}{\path|bodhan-ai/indic-transcribe-core|} \\
IndicConformer~\citep{indicconformer-monolingual} & \path|ai4bharat/indicconformer_stt_<lang>_hybrid_ctc_rnnt_large|; one monolingual checkpoint per language \\
Internal Hindi Parakeet & Internal checkpoint used for Hindi recovery ASR \\
\bottomrule
\end{tabularx}
\end{table}

\input{sections/appendix_prompts}

\section{Release Record Schema and Quality Rules}
\label{app:quality-rules}

\begin{table}[H]
\centering
\small
\caption{Fields in the released Granary-v2 manifests, in the order they appear on disk. Optional task labels can be empty when no usable target is available.}
\label{tab:label-schema}
\begin{tabularx}{\linewidth}{@{}p{0.47\linewidth}Y@{}}
\toprule
Fields & Meaning \\
\midrule
\path|audio_filepath| & Per-record identifier, not a resolvable path \\
\path|duration| & Audio duration in seconds \\
\path|corpus|, \path|source_lang| & Source corpus identifier and language code \\
\path|tn_raw| & Spoken-form transcript \\
\path|itn_raw|, \path|itn_clean| & Written forms with and without disfluencies \\
\path|captioning_text| & Short transcript summary \\
\path|code_switched_text| & Transcript with restored English spelling where applicable \\
\path|speech_qa_text| & Question--answer text, or \texttt{SKIP} \\
\path|context_asr| & Context object, or null \\
\path|translations| & Target-language translations, quality scores, and translation-stage notes \\
\path|preference_instructions| & List of instruction--target pairs and task tags \\
\path|high_quality| & Boolean flag for high-quality label selection, or null \\
\bottomrule
\end{tabularx}
\end{table}

The source audio used for labeling is 16~kHz mono. The released manifests contain audio identifiers rather than audio and carry no per-record sample-rate or channel-count field.

\paragraph{Nested objects.}
The \texttt{context\_asr} object contains coarse and fine context terms, entity categories, distractor terms, and five context prompts. Its confidence, speaking-style, and difficulty values are language-model estimates. The \texttt{translations} object is keyed by target-language code. Each translation contains \texttt{translation\_raw}, \texttt{translation\_quality\_score}, and \texttt{additional\_notes} with translation-stage status. Each \texttt{preference\_instructions} entry contains \texttt{prompt}, \texttt{target}, and \texttt{tags}. The \texttt{type} and \texttt{target\_lang} tags identify the task and requested target language.

\paragraph{Missing targets.}
Context can be null, and the translation object can be empty. SpeechQA can return \texttt{SKIP} when the transcript does not support a suitable question. The packer excludes empty values and skip responses from training pairs. The \texttt{high\_quality} flag supports filtering the released records. Intermediate ASR hypotheses and the full processing history are internal fields, separate from the release schema described here.

\paragraph{Character-scored languages.}
The CER set is Chinese, Japanese, Thai, Cantonese, Lao, Khmer, and Burmese. CER removes whitespace before comparison. Character-rate checks use the language-specific floors given in Section~\ref{sec:text-pipeline}.

\paragraph{Human-reference exceptions.}
Where no supported Qwen3-Omni improvement is available, the aligned human label can be retained. A further rescue rule applies to Ukrainian, Filipino, Maltese, Persian, Estonian, Dutch, Croatian, Lithuanian, Slovenian, Malayalam, Hebrew, Assamese, and Tamil. With a non-empty human reference, a record can be rescued when either ASR candidate has an error rate of at most 0.10 against that reference. Rescue also depends on the structural checks and active exceptions.

\paragraph{Language-pair exceptions.}
Declared and detected languages can differ for configured Croatian/Bosnian and Filipino/Tagalog pairs. The rule accepts the detected partner in the configured pair and rejects unrelated language mismatches.

\input{sections/appendix_label_evaluation}

\input{sections/appendix_training}

\input{sections/appendix_text_models}

%% file: tables/source_corpora.tex
\begin{table}[H]
\centering\scriptsize
\setlength{\tabcolsep}{2pt}
\caption{Corpus entries and durations for the full processed collection. The last column identifies public label availability; YODAS2, YouTube-Commons, and ReazonSpeech-v2 are excluded. Only LibriSpeech-PC is retained for LibriSpeech. Translation hours refer to audio already counted for ASR. Translation runs in both directions: non-English audio into English, and English audio into 31 target languages. A dash marks a direction that does not apply to the corpus. High-quality and translation-out hours follow the dataset card's integer rounding.}
\label{tab:source-corpora}
\begin{tabularx}{\linewidth}{@{}Yrrrrrc@{}}
\toprule
Corpus entry & Langs & ASR hours & \shortstack{High-quality\\hours} & \shortstack{X$\to$En\\hours} & \shortstack{En$\to$X\\hours} & \shortstack{Public\\release} \\
\midrule
VoxPopuli (MOSEL labels)~\citep{voxpopuli,mosel} & 23 & 380,711.26 & 123,192 & 84,999.29 & 8,990 & Included \\
YODAS2~\citep{yodas} & 22 & 222,913.89 & 95,574 & 35,517.18 & 5,848 & Excluded \\
YouTube-Commons~\citep{youtube-commons} & 24 & 122,258.88 & 45,354 & 3,261.71 & 10,549 & Excluded \\
LibriLight~\citep{librilight} & 1 & 49,334.35 & 25,898 & 0.00 & 25,725 & Included \\
ReazonSpeech-v2~\citep{reazonspeech-v2} & 1 & 35,344.08 & 11,504 & 11,503.98 & -- & Excluded \\
People's Speech (gov-origin subset)~\citep{peoples-speech} & 1 & 8,597.76 & 3,807 & 0.00 & 826 & Included \\
Multilingual LibriSpeech~\citep{mls} & 8 & 7,804.56 & 3,073 & 1,847.54 & 103 & Included \\
FalAR~\citep{falar} & 1 & 4,714.83 & 1,605 & 1,601.57 & -- & Included \\
Common Voice 26.0~\citep{commonvoice,commonvoice-v26} & 39 & 3,871.23 & 2,837 & 2,371.89 & 496 & Included \\
National Speech Corpus~\citep{national-speech-corpus} & 1 & 2,784.60 & 1,647 & 0.00 & 1,689 & Included \\
VoxPopuli (re-labelled)~\citep{voxpopuli} & 16 & 1,276.96 & 527 & 562.21 & 160 & Included \\
Golos~\citep{golos} & 1 & 1,207.72 & 715 & 743.57 & -- & Included \\
Europarl-ASR~\citep{europarl-asr} & 1 & 1,022.70 & 605 & 0.00 & 613 & Included \\
LibriSpeech-PC~\citep{librispeech-pc,panayotov2015librispeech} & 1 & 952.88 & 953 & 0.00 & 686 & Included \\
FLEURS~\citep{conneau2022fleurs} & 48 & 412.18 & 213 & 226.48 & 5 & Included \\
Dusha~\citep{dusha} & 1 & 203.73 & 104 & 143.01 & -- & Included \\
VCTK~\citep{vctk} & 1 & 82.15 & 76 & 0.00 & 76 & Included \\
AMI Meeting Corpus~\citep{carletta2005ami} & 1 & 70.38 & 24 & 0.00 & 4 & Included \\
Google Speech Commands~\citep{speech-commands} & 1 & 15.08 & 15 & 0.00 & 2 & Included \\
\midrule
Total & 48 & 843,579.22 & 317,724 & 142,778.43 & 55,772 & -- \\
\bottomrule
\end{tabularx}
\end{table}

%% file: sections/appendix_reference_config.tex
\subsection{Reference Settings}
\label{app:reference-config}

The metadata pipeline uses \path|nvidia/Nemotron-3-Diarization| with SpeechBrain language identification. The diarization model supports streaming and offline inference with up to eight speakers~\citep{nemotron3-diarization}. The settings below describe the reference metadata configuration for raw or long recordings. Granary-v2's short-form inputs bypass this stage.

The reference metadata launch uses VAD threshold 0.5, a 1,000~ms minimum interval, and 100~ms speech padding. Segments remain within 0.5--40 seconds. The diarization batch size is 2. Reader and writer concurrency are 3 and 4, respectively, with eight reader I/O threads.

\begin{table}[H]
\centering\small
\caption{Text-generation launch defaults in the reference configuration. The core-text column describes the Qwen configuration.}
\label{tab:reference-generation}
\begin{tabularx}{\linewidth}{@{}p{0.27\linewidth}YYY@{}}
\toprule
Setting & Core text stages & Contextual ASR & Translation \\
\midrule
Model & Qwen3.6-27B-FP8 & Qwen3.6-27B-FP8 & Qwen3.5-122B-A10B-FP8 \\
Allocated GPUs per node & 8 & 8 & 8 \\
Context / output limit & 2,048 / 512 tokens & 8,192 / 512 tokens & 1,024 / 256 tokens \\
Temperature / top-$p$ & 0 / 1 & 0.1 / 0.95 & 0.7 / 0.8 \\
Tensor parallelism & 2 & 2 & 2 per NMT replica \\
Replicas & 4 & 4 & 2 NMT replicas \\
Workers / batch & 12 / 128 & 8 / 64 & 2 / 256 \\
GPU memory fraction & 0.85 & 0.85 & 0.90 \\
\bottomrule
\end{tabularx}
\end{table}

The translation layout reserves eight GPUs and uses four for translation models and two for quality estimation. GPU-hour accounting includes all eight allocated GPUs. Its reference settings include top-$k$ 20, seed 1234, and a length-ratio limit of 8. Appendix~\ref{app:packing-details} gives the translation-pair selection threshold.

The official tutorial ASR example uses temperature 0, top-$k$ 1, repetition penalty 1, and seed 1234. It sets a 4,096-token context and a 256-token output limit.

\subsection{Text Validation and Fallback}
\label{app:validation-config}

The reference text stage checks for empty output, new Latin words outside its permitted set, and invalid word-count changes. Failed validation restores the input text and records a fallback note. These checks operate on vocabulary and word counts.

In the reference serial path, disfluency removal reads \texttt{itn\_raw}. Its word-count threshold is 0.5. For input of at least eight whitespace-separated tokens, output with fewer than half the input token count fails the deletion check. Other validation checks can also trigger fallback. The original written form is retained after a failed check.

\subsection{Compute Accounting Method}
\label{app:compute-method}

For job $j$, let $g_j$ be its fixed GPU allocation and $t_j$ its elapsed allocation time in hours. Logged compute is
\begin{equation}
G_{\mathrm{logged}}=\sum_j g_j t_j.
\end{equation}
The calculation includes startup, failed work, and retries within each allocation. Each allocation is counted once, including its nested job steps. CPU-node-hours are reported separately. Queue delays and job concurrency determine calendar time.

For work without a job record, compute can be estimated from the stage's input duration and processing rate. Context and translation estimates use the eligible records, with a separate workload for each translation direction. Pipeline processing, model training, and evaluation are accounted for separately.

%% file: sections/appendix_prompts.tex
\section{Label-Generation Prompts}
\label{app:prompts}

\begin{table}[H]
\centering\small
\caption{Prompt index. Label-generation prompts are separate from the training instructions in Appendix~\ref{app:packing-details}.}
\begin{tabularx}{\linewidth}{@{}Yl@{}}
\toprule
Task & Appendix reference \\
\midrule
Audio-only ASR & \ref{app:asr-prompts} \\
Human-reference ASR & \ref{app:reference-prompts} \\
Entity recovery and punctuation & \ref{app:er-pnc-prompts} \\
TN, language ID, ITN, disfluency removal & \ref{app:normalization-prompts} \\
Captioning, code switching, SpeechQA & \ref{app:auxiliary-prompts} \\
Context extraction & \ref{app:context-prompts} \\
Text translation & \ref{app:translation-prompts} \\
\bottomrule
\end{tabularx}
\end{table}

The listings reproduce the archived source prompts. Braced terms, such as
\texttt{\{text\}}, are substitution fields for the record's text or language
name. The \texttt{SYSTEM\_PROMPT} and \texttt{USER\_PROMPT\_TEMPLATE} markers
identify the system and user roles.

The source bundle contains the original UTF-8 prompt files and their hashes.
The listings show non-ASCII code points as Unicode escapes, such as
\texttt{\textbackslash u2014}. Model inputs use the original UTF-8 files.

\subsection{Audio-Only ASR}
\label{app:asr-prompts}
The official tutorial reference provides separate English and multilingual prompts. Both are user
instructions followed by the audio, using the model's chat template.
\par\medskip\noindent\textbf{English.}\par\nopagebreak[4]\smallskip
\begin{lstlisting}[style=granaryprompt]
You receive English audio. The audio is the ground truth.

MAIN GOAL: Listen carefully to the audio and transcribe exactly what is spoken, including all disfluencies present in the audio.
- Do NOT invent words or content not spoken in the audio.
- Do NOT remove substantive content that is spoken in the audio.
- Do NOT paraphrase, polish grammar or rewrite sentences.

ENTITIES (names, places, brands, titles, etc.):
- Keep every named entity as spoken. This includes names, places, brands, titles, acronyms, and other proper nouns.
- Do not ever transliterate, translate, re-spell, normalize, or "correct" an entity into another script.
- If enetities are part code switched data it should stay the same.

KEEP DISFLUENCIES:
- Keep fillers, repetitions, false starts, colloquial reductions, and grammatical errors spoken in the audio.
- Add hesitation markers and fillers natural to English wherever they are spoken in the audio.
- Do NOT clean up, normalize, or remove disfluencies spoken in the audio.
- Add consecutive instances of the same word or short phrase when spoken unintentionally.
  - Example: "I I think" if that is what is spoken.

BACKGROUND / QUIET / OVERLAPPING SPEECH:
- Keep all audible speech, including quieter, distant, or overlapping voices \u2014 not just the loudest speaker.
- Add background or secondary speech that is audible; do not drop words because they sound like background.

FALSE STARTS:
- Add incomplete words or phrases the speaker abandons, marked with a hyphen.
  - Example: "I was go- going to the store."
- Do NOT remove false starts spoken in the audio.

COLLOQUIAL REDUCTIONS:
- If the speaker used reductions, use the spoken form: "want to" \u2192 "wanna", "going to" \u2192 "gonna", etc.
- Preserve forms such as "wanna", "gonna", "kinda", "lemme", "lotta", "outta", "Imma", "sorta", "ya", "m'kay", "finna", "tryna", etc. Do NOT expand them.

WRONG GRAMMAR:
- Keep grammatical errors as spoken. Do NOT correct subject-verb agreement, tense errors, or other grammar issues.

NUMERICALS:
- Keep numbers as spoken in words. Do NOT convert them to digits.
  - Example: keep "oh eleven" or "zero eleven".

Output format:
- Return ONLY the transcription text.
\end{lstlisting}

\par\medskip\noindent\textbf{Multilingual.}\par\nopagebreak[4]\smallskip
\begin{lstlisting}[style=granaryprompt]
You receive an audio file.

Goal: To transcribe the audio, normalize numbers into words, and include any disfluencies that are present in the audio.

TRANSCRIPTION RULES:
1) Normalize numeric expressions into words exactly as they are SPOKEN in the audio.
- Mixed format is forbidden:
    Bad: "5 percent", "2 zeros"
    Good: "five percent", "two zeros"
- Normalize: percentages, currencies, units, ranges, decimals, dates/years \u2014 ONLY if they are spoken.
- If a unit (for example \u201cpercent\u201d) is NOT spoken, do not add it.
2) Include any disfluencies present in the audio.
- Disfluencies as "um", "uh" that are present in the audio should be added to the text.
- If a word is repeated in the audio, include every repetition in the text.

ENTITIES (names, places, brands, titles, etc.):
- Keep every named entity as spoken. This includes names, places, brands, titles, acronyms, and other proper nouns.

OUTPUT FORMAT:
- Return only the final text.
- No explanations, no JSON, no lists.
\end{lstlisting}

\subsection{ASR with a Human Reference}
\label{app:reference-prompts}
The archived reference-label prompt supplies the source transcript in \texttt{\{transcript\}}. The audio remains
the source for spoken content. Entity spellings can come from the reference.
\par\medskip\noindent\textbf{English reference improvement.}\par\nopagebreak[4]\smallskip
\begin{lstlisting}[style=granaryprompt]
You receive English audio and a reference transcript. The reference may be cleaned, partially wrong, or missing speech artifacts. The audio is the ground truth.

REFERENCE TRANSCRIPT:
"{transcript}"

MAIN GOAL: Listen carefully to the audio and revise the reference so it faithfully reflects exactly what is spoken, including all disfluencies present in the audio.
- Use the reference as a starting point; do not ignore it.
- When the reference matches the audio, keep it unchanged.
- When the reference conflicts with the audio, follow the audio.
- Do NOT invent words or content not spoken in the audio.
- Do NOT remove substantive content that is spoken in the audio (remove reference words only if they are not spoken).
- Do NOT paraphrase, polish grammar or rewrite sentences that already match the audio.
- Prefer minimal edits: fix mismatches and insert missing speech artifacts.
- Preserve named entities from the reference in their exact written form.
- Normalize numbers to their written form.

ENTITIES (names, places, brands, titles, etc.):
- Keep every named entity from the reference in its exact written form: spelling, casing, script, and punctuation. This includes names, places, brands, titles, acronyms, and other proper nouns.
- Do not ever transliterate, translate, re-spell, normalize, or "correct" an entity into another script.
- If enetities are part code switched data it should stay the same.

KEEP REFERENCE DISFLUENCIES:
- If the reference already has fillers, repetitions, false starts, colloquial reductions, or grammatical errors, keep them.
- Add hesitation markers and fillers natural to English wherever they are spoken in the audio but missing from the reference.
- Do NOT clean up, normalize, or remove disfluencies that are already in the reference and are spoken in the audio.
- Add consecutive instances of the same word or short phrase when spoken unintentionally.
  - Example: reference "I think" \u2192 "I I think" if that is what is spoken.

BACKGROUND / QUIET / OVERLAPPING SPEECH:
- Keep all audible speech in the reference, including quieter, distant, or overlapping voices \u2014 not just the loudest speaker.
- Add background or secondary speech that is audible but missing; do not drop words because they sound like background.

FALSE STARTS:
- Add incomplete words or phrases the speaker abandons, marked with a hyphen.
  - Example: "I was go- going to the store."
- Do NOT remove false starts already in the reference if they are spoken in the audio.

COLLOQUIAL REDUCTIONS:
- If the reference uses standard forms but the speaker used reductions, use the spoken form: "want to" \u2192 "wanna", "going to" \u2192 "gonna", etc.
- Preserve forms such as "wanna", "gonna", "kinda", "lemme", "lotta", "outta", "Imma", "sorta", "ya", "m'kay", "finna", "tryna", etc. Do NOT expand them.

WRONG GRAMMAR:
- Keep grammatical errors as spoken. Do NOT correct subject-verb agreement, tense errors, or other grammar issues.

NUMERICALS:
- Keep numbers as spoken in words. Do NOT convert them to digits.
  - Example: keep "oh eleven" or "zero eleven".

Output format:
- Return ONLY the revised transcription text.
\end{lstlisting}

The archived English reference prompt contains conflicting instructions for written and spoken number forms.

\par\medskip\noindent\textbf{Multilingual reference improvement.}\par\nopagebreak[4]\smallskip
\begin{lstlisting}[style=granaryprompt]
You receive:
1) An audio file,
2) A Ground Truth Transcription of the audio {transcript}.

Goal: To normalize numbers from the text and add any disfluencies that are present in the audio.

ALLOWED ONLY:
1) Normalize numeric expressions into words exactly as they are SPOKEN in the audio.
- Mixed format is forbidden:
    Bad: "5 percent", "2 zeros"
    Good: "five percent", "two zeros"
- Normalize: percentages, currencies, units, ranges, decimals, dates/years \u2014 ONLY if they are spoken.
- If a unit (for example \u201cpercent\u201d) is NOT spoken, do not add it.
2) Add any disfluencies present in the audio.
- Disfluencies as "um", "uh" that are present in the audio should be added to the text.
- If word is repeated in the audio but missing from ground truth add it to the text.

ENTITIES (names, places, brands, titles, etc.) should be the same as inGround Truth Transcription:
- Keep every named entity from the reference in its exact written form: spelling, casing, script, and punctuation. This includes names, places, brands, titles, acronyms, and other proper nouns.

OUTPUT FORMAT:
- Return only the final text.
- No explanations, no JSON, no lists.
\end{lstlisting}

\subsection{Entity Recovery and Punctuation}
\label{app:er-pnc-prompts}
Entity recovery uses a system instruction and a user template with two text
fields. It runs only where a trusted human reference is available.
\par\medskip\noindent\textbf{Entity recovery.}\par\nopagebreak[4]\smallskip
\begin{lstlisting}[style=granaryprompt]
# SYSTEM_PROMPT

You are given TWO transcriptions of the same audio clip:

- INITIAL: Initial reference with named entity spelling/casing.
- NORMALIZED: Normalized transcription.

Your task:

1. In the INITIAL text, identify ONLY proper-noun named entities made of letters/words only (no digits): people, places, organizations, companies, products, brands, and acronyms/initialisms. 

2. If a candidate contains any digit (0-9), it is NOT an entity and must be excluded.

3. Titles/honorifics are NOT entities and must be excluded.

4. For each remaining entity, locate its match in the NORMALIZED text and substitute it with the INITIAL spelling/casing. Change ONLY the entity itself \u2014 leave all normalization untouched: keep numbers, dates, times, money, units, and titles in their written form.

Return ONLY the resulting NORMALIZED text. No explanations, labels, or extra formatting.

# USER_PROMPT_TEMPLATE

INITIAL:
"{ground_truth}"

NORMALIZED:
"{normalized}"

\end{lstlisting}

\par\medskip\noindent\textbf{General punctuation and capitalization.}\par\nopagebreak[4]\smallskip
\begin{lstlisting}[style=granaryprompt]
Add punctuation and capitalization to the raw ASR transcription in {language} below. That is the ONLY change you are allowed to make.
Raw ASR transcription in {language}:
{text}

RULES:
- Add periods, commas, question marks, exclamation marks where appropriate.
- Capitalize sentence starts and proper nouns only.
- Every single word from the input MUST appear in the output, unchanged and in the same order. Do not remove, add, replace, or reorder any words or characters.

Return ONLY the punctuated and capitalized transcription. No explanations.
\end{lstlisting}

\subsection{Normalization and Transcript Language}
\label{app:normalization-prompts}
\par\medskip\noindent\textbf{Text normalization.}\par\nopagebreak[4]\smallskip
This prompt maps written expressions to spoken forms. It preserves non-target
words and speech disfluencies.
\begin{lstlisting}[style=granaryprompt]
# Text Normalization ({language})

Convert the written-form transcript below in **{language}** to spoken form. **Only** apply conversions from the table (numbers, dates, times, money, units, symbols, etc.). Use **{language}** conventions; English examples show the pattern.

Return ONLY normalized text. No explanations or extra formatting.

## Constraints

- Stay in **{language}**; preserve code-switching/script mix. Do NOT translate.
- Do NOT clean up speech: keep fillers (um, uh, \u2026), repetitions, false starts (go- going), colloquial forms (gonna, wanna), grammar errors, and all non-target wording unchanged and in order.
- Do NOT add or remove words or punctuation except as required by conversions (e.g. `@` \u2192 "at" in an email).
- Denormalize digits/symbols **only** when they match Conversion Rules. Leave idiomatic numbers, proper nouns, vague quantities, disfluencies, and text already spoken as words unchanged (e.g. keep "oh eleven" as-is; do not swap "oh" \u2194 "zero" unless denormalizing a digit sequence).
- In stammers/false starts, denormalize only the final clean numeric token.

## Conversion Rules

| Category | Written | Spoken |
|---|---|---|
| Cardinal | 14 / 1,030.5 / 2024 | fourteen / one thousand thirty point five / twenty twenty four |
| Ordinal | 1st / 21st / 50th | first / twenty first / fiftieth |
| Date | January 22nd, 1847 / January 22, 1990 | january twenty second eighteen forty seven / january twenty two nineteen ninety |
| Time | 3:05 PM / 10 AM / 1:45 / noon | three o five PM / ten AM / one forty five / noon |
| Money | $52 / $249.99 | fifty two dollars / two hundred forty nine dollars and ninety nine cents |
| Percent | 0.5% / 20% to 30% | zero point five percent / twenty to thirty percent |
| Units | 5 kg / 90 km/h / 5'4" | five kilograms / ninety kilometers per hour / five foot four |
| Fractions | 1/2 / 1/3 / 2/3 / 1 3/4 | half / a third / two thirds / one and three quarters |
| Phone | 5558675309 / 18005550199 | five five five eight six seven five three zero nine / one eight hundred five five five oh one nine nine |
| URL/Email | example.com/pricing / john@gmail.com | example dot com slash pricing / john at gmail dot com |
| Titles | Dr. Smith / Prof. Jones / vs. / w/o | doctor smith / professor jones / versus / without |
| Address | 450 N. Main St. | four fifty north main street |
| Roman num. | King Henry VIII / Chapter IV | king henry the eighth / chapter four |
| Negative | -12 / -5 degrees | negative twelve / minus five degrees |
| Decades | 70s / 20s | seventies / twenties |
| Letter+num | Q2 / B12 | q two / b twelve |

Additional rules:
- Structural symbols (`.`, `@`, `/`, `:`, `-`) \u2192 spoken words in URLs, emails, phones, using **{language}** forms ("dot", "at", "slash", \u2026).
- Acronyms (NASA, FBI, AM, PM): keep natural spoken form; do not expand unrelated abbreviations.
- Phone/time "0" \u2192 "oh" or "zero" per **{language}**; zip/house numbers \u2192 digit-by-digit when denormalizing.

## Ambiguity

- Denormalize digit/symbol form for quantities, measurements, ages, dates, and counts.
- Skip idioms, proper nouns ("One Direction"), vague quantities, temporal "quarter", and idiomatic "half".
- Convert 1/2 or 1/4 to "half" / "a quarter" only for true fractions.

Written-form transcript in {language}:
{text}
\end{lstlisting}

\par\medskip\noindent\textbf{Transcript-language identification.}\par\nopagebreak[4]\smallskip
The model returns a primary language name and the languages present in the
text. Verification then resolves names and codes for comparison with
the source language. This text check is separate from acoustic language ID.
\begin{lstlisting}[style=granaryprompt]
# Language Identification

You are given a transcript of spoken audio. Your task is to identify the language(s) in which the text is written, including any code-switching.

Analyze the vocabulary, grammar, script, and word patterns. Spoken audio can be code-switched as well \u2014 a speaker may mix two or more languages in a single utterance. Detect this instead of collapsing everything to one label.

## Guidelines

- Identify the PRIMARY language: the one that carries most of the meaningful content.
- Also list EVERY language present in the transcript, including the primary one.
- Return full English names of languages (e.g. English, Spanish, French, German, Hindi, Chinese). Do not return ISO codes, locale tags, or abbreviations (e.g. do not return "en", "es", or "fr-FR").
- Proper nouns and isolated loanwords do NOT count as a separate language; only count a language if it contributes real words or phrases.
- If only one language is present, list just that one.
- If the text is too short to identify confidently, choose the most likely language(s) based on the available words and script.

## Output format

Return exactly two lines, nothing else. No explanations, quotes, or extra formatting:

```
Primary: <primary language name>
Languages: <comma-separated language names, including the primary>
```

## Examples

Transcript: "Bonjour, comment allez-vous aujourd'hui?"
Primary: French
Languages: French

Transcript: "El clima est\u00e1 muy agradable esta ma\u00f1ana."
Primary: Spanish
Languages: Spanish

Transcript: "\u092e\u0941\u091d\u0947 \u0932\u0917\u0924\u093e \u0939\u0948 \u0915\u093f \u0939\u092e\u0947\u0902 \u0907\u0938 project \u0915\u094b \u0915\u0932 \u0924\u0915 finish \u0915\u0930 \u0926\u0947\u0928\u093e \u091a\u093e\u0939\u093f\u090f\u0964"
Primary: Hindi
Languages: Hindi, English

Transcript: "I think we should schedule the meeting for next Tuesday."
Primary: English
Languages: English
\end{lstlisting}

\par\medskip\noindent\textbf{Inverse text normalization.}\par\nopagebreak[4]\smallskip
\begin{lstlisting}[style=granaryprompt]
# Inverse Text Normalization

Convert spoken-form text to standard written form: numbers, dates, times, currencies, measurements, and symbols become their conventional written representations.

Return ONLY the converted text. No explanations, labels, or extra formatting.

## Constraints

- PRESERVE the input language. Do NOT translate.
- PRESERVE all original wording, disfluencies (um, uh, euh, \u00e4hm), repetitions, false starts (go- going), colloquial forms (gonna, 'cause), mispronunciations, and grammatical errors.
- Do NOT add punctuation that wasn't implied by the input.
- Do NOT paraphrase, add, or remove words beyond the conversions below.
- When a number word is idiomatic (not numeric), keep it as a word (see Ambiguity Resolution).

## Conversion Rules

| Category | Spoken | Written |
|---|---|---|
| Cardinal | fourteen / one thousand thirty point five / twenty twenty four | 14 / 1,030.5 / 2024 |
| Ordinal | first / twenty first / fiftieth | 1st / 21st / 50th |
| Date | january twenty second eighteen forty seven / january twenty two nineteen ninety | January 22nd, 1847 / January 22, 1990 |
| Time | three o five PM / ten AM / one forty five / noon | 3:05 PM / 10 AM / 1:45 / noon |
| Money | fifty two dollars / two hundred forty nine dollars and ninety nine cents | $52 / $249.99 |
| Percent | zero point five percent / twenty to thirty percent | 0.5% / 20% to 30% |
| Units | five kilograms / ninety kilometers per hour / five foot four | 5 kg / 90 km/h / 5'4" |
| Fractions | half / a third / two thirds / one and three quarters | 1/2 / 1/3 / 2/3 / 1 3/4 |
| Phone | five five five eight six seven five three zero nine / one eight hundred five five five oh one nine nine | 5558675309 / 18005550199 |
| URL/Email | example dot com slash pricing / john at gmail dot com | example.com/pricing / john@gmail.com |
| Titles | doctor smith / professor jones / versus / without | Dr. Smith / Prof. Jones / vs. / w/o |
| Address | four fifty north main street | 450 N. Main St. |
| Roman num. | king henry the eighth / chapter four | King Henry VIII / Chapter IV |
| Negative | negative twelve / minus five degrees | -12 / -5 degrees |
| Decades | seventies / twenties | 70s / 20s |
| Letter+num | q two / b twelve | Q2 / B12 |

Additional rules:
- Symbols: "dot"\u2192".", "at"\u2192"@", "slash"\u2192"/", "colon"\u2192":", "dash"\u2192"-".
- Acronyms (NASA, FBI, SQL, AM, PM): keep uppercase as-is.
- "zero" \u2192 "0"; "oh"/"o" in phone/time contexts \u2192 "0".
- Zip codes: spell digit by digit. House numbers: digits.

## Ambiguity Resolution

- Prefer digits for quantities, measurements, ages, dates, counts.
- Keep as WORDS for idioms, proper nouns, and pronominal/indefinite uses:
  - "one of the best", "the only one", "one another", "one by one"
  - "One Direction" (proper noun), "a couple of things" (vague quantity)
- "quarter" in temporal/financial contexts stays as "quarter" ("fourth quarter", "quarter over quarter"). Only convert to 1/4 when it's a true fraction ("a quarter of a cup" \u2192 "1/4 of a cup").
- "half" \u2192 1/2 only as a fraction; idiomatic use stays a word ("half the time", "half asleep").
- In stammers/false starts, keep number words as words: "o- o- one hundred" \u2192 "o- o- 100" (digits only on the final clean token).
\end{lstlisting}

\par\medskip\noindent\textbf{Disfluency removal.}\par\nopagebreak[4]\smallskip
The serial reference path applies this prompt to \texttt{itn\_raw} after ITN
to remove disfluencies from the written-form transcript.
\begin{lstlisting}[style=granaryprompt]
# Disfluency Removal

Remove speech disfluencies from the following text. Do NOT change anything else.

Return ONLY the cleaned text. No explanations, labels, or extra formatting.

## What to Remove

- **Filler words**: "um", "uh", "hm", "ah", "er", "eh", "like" (when used as filler), "you know" (when used as filler)
- **Repeated words**: "I I think" \u2192 "I think", "the the problem" \u2192 "the problem"
- **False starts**: incomplete words the speaker abandoned, e.g. "I was go- going to" \u2192 "I was going to"

## What to Keep \u2014 Do NOT Change

- All numbers, dates, symbols, abbreviations \u2014 leave exactly as they are
- Punctuation and capitalization \u2014 leave exactly as they are
- Colloquial forms ("gonna", "wanna", "'cause") \u2014 keep as-is
- Grammar \u2014 do NOT correct grammatical errors
- Content words \u2014 do NOT rephrase or paraphrase
- The language \u2014 do NOT translate

The ONLY change you should make is removing filler words, repeated words, and false starts. Everything else must remain identical.
\end{lstlisting}

\subsection{Captioning, Code Switching, and SpeechQA}
\label{app:auxiliary-prompts}
These prompts use the speech-faithful transcript as their input.
A caption describes transcript content. A SpeechQA answer must be supported
by that content. The exact \texttt{SKIP} response is an allowed task outcome. It is excluded from question--answer training pairs.
\par\medskip\noindent\textbf{Captioning.}\par\nopagebreak[4]\smallskip
\begin{lstlisting}[style=granaryprompt]
# Audio Caption Generation

You are given a transcript of spoken audio. Generate a short, descriptive caption that summarizes what is being said.

Return ONLY the caption. No explanations, labels, quotes, or extra formatting.

## Guidelines

- Write a concise summary (1-2 sentences) of the main topic or content
- Use third person (e.g. "A speaker discusses..." or "The conversation covers...")
- Capture the key subject matter, not every detail
- Keep it factual \u2014 do not add opinions or interpretations
- If the transcript is too short or unclear, describe what little content there is
- PRESERVE the language of the input \u2014 if the transcript is in Spanish, write the caption in Spanish

## Examples

Transcript: "So today we're going to talk about how photosynthesis works in plants. The basic idea is that plants convert sunlight into chemical energy."
Caption: A speaker explains how photosynthesis works, describing how plants convert sunlight into chemical energy.

Transcript: "The stock market dropped three percent yesterday following the Federal Reserve's announcement about interest rates."
Caption: A news report about a stock market decline following a Federal Reserve interest rate announcement.

Transcript: "First you need to preheat the oven to three fifty. Then mix the flour and sugar together in a large bowl."
Caption: Instructions for a baking recipe, including oven temperature and mixing dry ingredients.
\end{lstlisting}

\par\medskip\noindent\textbf{Code-switching restoration.}\par\nopagebreak[4]\smallskip
\label{app:code-switching-prompt}
This prompt reads \texttt{tn\_raw} and writes
\texttt{code\_switched\_text}. It preserves the native-language words and
changes supported English-origin forms to Latin spelling. It also preserves
native-script suffixes attached to restored English roots. The model retains
the input when no restoration is needed. Appendix~\ref{app:packing-details} shows how
this output becomes a separate training target.
\begin{lstlisting}[style=granaryprompt]
You are a {language} language expert. Your task is to restore code-switching in a {language} transcript: whenever a word in the transcript is actually an English word that was written phonetically in {language} script, rewrite it using its original English (Latin) spelling.

Context: the input transcript was produced by a human transcriber who wrote everything in {language} script, including English words the speaker actually pronounced in English. We want the transcript to reflect the real code-switched utterance \u2014 {language} words stay in {language} script, English words appear in Latin script.

You are performing transliteration RESTORATION, NOT translation. This distinction is the entire task.

THE PHONETIC TEST (apply to every single word):
For each word written in {language} script, ask: "If I pronounce this word using {language} phonology, does the resulting sound resemble an English word with the same meaning?"
- If YES \u2014 the word is an English loanword spelled out phonetically in {language} script. Restore its standard English Latin spelling.
- If NO \u2014 the word is a native {language} word. LEAVE IT EXACTLY AS WRITTEN. Do not translate it, even if you know what it means in English.

The phonetic test is about sound, not meaning. A {language}-native word can mean the same thing as an English word and still be native. Native vocabulary in any language has its own phonology that does not echo English. Loanwords inherit English phonemes that show through the transliteration.

WHEN IN DOUBT, LEAVE THE WORD UNCHANGED. Under-restoration is acceptable; over-translation is not. If you cannot clearly hear the English pronunciation in the {language}-script word, treat it as native.

Categories that ARE typically loanwords (restore to English):
- Modern technology terms, scientific/medical/chemical terminology
- Brand names, company names, product names
- Person names of non-{language} origin, place names of non-{language} origin
- Acronyms and abbreviations
- Recently borrowed concepts the language does not have indigenous words for

Categories that are typically NATIVE (leave unchanged):
- Everyday objects, common materials, basic concepts, body parts, family terms
- Native flora, fauna, foods, cultural/religious terms
- Place names native to the {language}'s region
- Words for which the {language} has used its own term for centuries

Grammatical rules:
1. If an English root carries a {language} grammatical suffix (plural, case marker, etc.), output the English root in its standard Latin spelling and keep the suffix attached in its original {language} script. Do NOT insert a space, hyphen, or any separator between the Latin root and the {language}-script suffix, and do NOT transliterate the suffix into Latin script.
2. Preserve original word order, punctuation, and sentence structure exactly. Do not paraphrase, summarize, or reorder.
3. If the input contains no English-origin transliterated words, return it unchanged.
4. Output ONLY the rewritten {language} text. No explanations, no quotation marks around the output, no commentary.

Apply the phonetic test to every word. When in doubt, leave the word unchanged. Now rewrite the following {language} text:
{text}
\end{lstlisting}

\noindent\begin{minipage}{\linewidth}
\par\medskip\noindent\textbf{SpeechQA.}\par\nopagebreak[4]\smallskip
\begin{lstlisting}[style=granaryprompt]
You are given a transcript of a spoken audio segment in {language}. Generate at most one question-answer (Q/A) pair grounded in the transcript content.

The Q/A pair must be answerable from the transcript alone \u2014 no external knowledge, no information that is not stated or directly implied in the transcript. The question and the answer must both be in {language}, the same language as the transcript.

First decide whether the transcript is suitable for Q/A generation. The transcript is NOT suitable if:
- It is too short or carries no substantive informational content (a few words, pure filler, disfluencies only, isolated single words).
- It is a pure conversational utterance with no information content (greetings, farewells, acknowledgments, agreements, fillers like "yeah", "okay", "thanks").

For everything else (borderline cases, edge cases, content of any genre or style), use your own judgment based on whether a meaningful, well-grounded Q/A pair can plausibly be produced. When borderline, prefer SKIP over generating a weak or hallucinated pair.

Output format:

If the transcript IS suitable, output exactly two lines and nothing else:
Q: <one concrete question in {language}, not yes/no, not a trivial restatement>
A: <a concise answer in {language}, a phrase or one short sentence, directly supported by the transcript>

If the transcript is NOT suitable, output exactly the single line:
SKIP

Do not output explanations, markdown, code fences, surrounding text, blank lines before, or anything other than the two-line Q/A block or the single word SKIP.

Transcript:
{text}
\end{lstlisting}
\end{minipage}

\subsection{Contextual ASR Extraction}
\label{app:context-prompts}
The reference context stage reads the curated transcript and source language.
It emits the JSON object specified below. A CPU step can then add acoustic
distractors and form the training instructions in Appendix~\ref{app:packing-details}.
Confidence and difficulty values in this response are language-model estimates.
\begin{lstlisting}[style=granaryprompt]
# SYSTEM_PROMPT

You are an expert ASR data annotator. Given a transcript of an audio clip in
the language specified by the user, extract contextual-ASR biasing information
that a downstream ASR model can use as context hints. Preserve every entity
name in its original form exactly as it appears in the transcript.

Return ONE JSON object with this shape (all 9 entity_categories keys must be
present, possibly empty):

```json
{
  "coarse_context_terms": [],
  "fine_context_terms": [],
  "entity_categories": {
    "person_name": [], "company_name": [], "product_name": [],
    "drug_name": [], "location_name": [], "organization_name": [],
    "event_name": [], "technical_term": [], "abbreviation": []
  },
  "distractor_terms": [],
  "confidence_coarse": 1, "confidence_fine": 1,
  "speaking_style": "conversational",
  "estimated_difficulty": 1
}
```

## Field rules

**coarse_context_terms** (1\u20133 items): always pick the MOST SPECIFIC domain
label you can justify from the transcript. Use `"Interventional Cardiology"`,
not `"Medicine"`; `"Quarterly Earnings Call"`, not `"Business"`;
`"Philosophy of Technology"`, not `"General Knowledge"`. Only fall back to
`"Daily Conversation"` / `"General Knowledge"` when the content is truly
everyday speech with no identifiable subject area, and set
`confidence_coarse` low in that case.

**fine_context_terms**: flat list. MUST equal the union of the 9
`entity_categories` lists. Use the exact form from the transcript \u2014 no
re-normalising case, spelling, or spacing.

**entity_categories** \u2014 9 strict buckets. Each entity goes in exactly ONE.
Capitalised \u2260 named entity. When in doubt, leave the bucket empty.

| Bucket | Belongs here | Counter-example (skip) |
|---|---|---|
| `person_name` | individuals, titled names: `Dr. Patel`, `Satya Nadella` | `the CEO`, `my manager` |
| `company_name` | for-profit companies: `NVIDIA`, `Sony`, `CD Projekt` | `the company` |
| `product_name` | branded products / software / services: `iPhone`, `ChatGPT`, `PlayStation` | `laptop`, `phone` |
| `drug_name` | drugs / compounds: `sitagliptin`, `Aspirin`, `Ozempic` | `painkiller` |
| `location_name` | cities / countries / landmarks: `San Francisco`, `China` | `the office` |
| `organization_name` | non-commercial orgs / agencies / universities: `NASA`, `WHO`, `Stanford University` | `the government` |
| `event_name` | events / conferences / occurrences: `Olympics`, `COVID`, `CES` | `the meeting` |
| `technical_term` | rare specialised non-proper-noun tokens: `backpropagation`, `arthroscopic`, `sitagliptin` | `system`, `synergy` |
| `abbreviation` | non-named-entity acronyms: `MRI`, `USB`, `API`, `GPU` | `ESG`, `EBITDA`, `KPI` (jargon \u2192 exclude entirely) |

Named-entity acronyms (`NASA`, `NVIDIA`, `COVID`) go in their entity bucket,
NOT `abbreviation`. Industry-jargon acronyms are excluded from all buckets.

**distractor_terms** (3\u20138 items): plausible same-domain entities NOT present
in the transcript. Must be real named entities or real domain terms \u2014 never
generic phrases. Used to teach the ASR model not to copy hints blindly.

**confidence_coarse / confidence_fine** (1\u20135): 5 = clear and unambiguous; 3 =
some uncertainty / multiple plausible interpretations; 1 = ambiguous, generic,
too short, or noisy.

**speaking_style**: pick one of `formal | conversational | technical | narrative | instructional` based on register.

**estimated_difficulty** (1\u20135): 5 = many rare words / spelled-out acronyms /
dense numerics; 3 = moderate domain vocabulary; 1 = everyday speech.

## Output

Return ONLY the JSON object. No markdown fences, no commentary. All 9
`entity_categories` keys present. `fine_context_terms` = union of those 9
lists. Do not invent entities not supported by the transcript.

# USER_PROMPT_TEMPLATE

Source language: {source_lang}

Transcript:
"{transcript}"

Extract contextual-ASR biasing information per the schema above and return ONLY the JSON object.
\end{lstlisting}

\subsection{Text Translation}
\label{app:translation-prompts}
Translation reads \texttt{tn\_raw}. The caller supplies full language names
for \path|{source_lang}| and \path|{target_lang}|. This is a
text-label-generation prompt, separate from the speech translation
instruction presented to a trained speech model.
\par\medskip\noindent\textbf{System turn.}\par\nopagebreak[4]\smallskip
\begin{lstlisting}[style=granaryprompt]
You are a professional translator.
  - Output only the translation.
  - The input may be incomplete. Preserve incompleteness.
\end{lstlisting}

\par\medskip\noindent\textbf{User turn.}\par\nopagebreak[4]\smallskip
\begin{lstlisting}[style=granaryprompt]
Translate the following {source_lang} segment into {target_lang}, without additional explanation.

{text}
\end{lstlisting}

The translation prompt requests preservation of incomplete input.
The translation-quality criteria in Table~\ref{tab:evaluation-dimensions}
also cover fillers, repetitions, and false starts.

%% file: sections/appendix_label_evaluation.tex
\subsection{Label Quality Checks}
\label{app:label-evaluation}

\input{tables/evaluation_dimensions}

The label-quality protocol uses human-reference subsets to compare primary and selected ASR outputs. It also compares single-model and recovery-model selection, source and improved human labels, and outputs before and after entity recovery. Results are grouped by script, language, corpus, and filtering rule to identify sources of error.

Targeted checks cover character-rate rules, WER and CER selection, human-reference exceptions, and language-pair exceptions. Human assessment covers the relevance of generated context terms and distractors. Translation-quality criteria include meaning, disfluencies, and incomplete spoken content.

%% file: tables/evaluation_dimensions.tex
\begin{table}[H]
\centering
  \caption{Evaluation dimensions for speech-faithful labels and translation.}
  \label{tab:evaluation-dimensions}
  \centering
  \small
  \begin{tabularx}{\linewidth}{@{}p{0.27\linewidth}Y@{}}
    \toprule
    Behavior & Example measure \\
    \midrule
    Speech-faithful output & WER or CER against a reference that preserves spoken forms and disfluencies \\
    Text normalization & Accuracy for numbers, dates, money, units, symbols, and written-to-spoken conversions \\
    Disfluency preservation & Precision and recall for fillers, repetitions, and false starts present in the audio \\
    Code-switched text & English spelling and script accuracy; native-word and suffix preservation; unsupported changes \\
    Output consistency & Rate of format violations and inconsistent normalization choices \\
    Translation fidelity & Semantic adequacy plus preservation of disfluencies and spoken content \\
    \bottomrule
  \end{tabularx}
\end{table}

%% file: sections/appendix_training.tex
\section{Reference Training Recipes and Packing}
\label{app:training-recipes}

The Transformer-TDT encoder comprises 32 layers with a hidden
dimension of 1280 and 10 attention heads, totaling 630.9M encoder
parameters. The English and multilingual configurations share
this architecture and use \texttt{tn\_raw} as the training target.
The English tokenizer contains 1,024 BPE units, while the
multilingual tokenizer contains 16,384. English training uses
indexed inputs; the multilingual configuration uses streaming
inputs without indexes.

Audio is sampled at 16\,kHz and represented by 128 log-mel
features computed with a 25\,ms window and a 10\,ms stride.
The encoder uses $8\times$ subsampling and rotary position
embeddings. The TDT duration set is $\{0,1,2,3,4\}$.
The optimization settings and training duration for both English
runs in Table~\ref{tab:english-quality-results} are described
in Section~\ref{sec:tdt-training}.

\subsection{Packing Rules and Context Sampling}
\label{app:packing-details}

Each \texttt{preference\_instructions} entry has \texttt{prompt}, \texttt{target}, and \texttt{tags}. Task and target-language tags describe the pair. The reference wrapper uses seed 42 and can join translations by stable shard path and audio identifier. It selects translations with a quality-estimation score of at least 0.7 for training pairs. The translation manifest retains the scored outputs.

The packer prefers \texttt{tn\_raw} for transcription and uses \texttt{pnc\_text} when the spoken-form field is absent. Prompt selection also uses the recorded punctuation-stage status. Written-form instructions use \texttt{itn\_raw}; disfluency-free instructions use \texttt{itn\_clean}. SpeechQA parses the generated question into the instruction and the answer into the target. Empty text, invalid target-language codes, negative sentinel scores, and skip responses are not valid training targets.

Partial-context sampling retains a fraction between 0.5 and 0.8 of the relevant terms. The reference seed is 42. Phonetic matching can retain up to three candidates per entity, eight acoustic distractors, and sixteen combined distractors. Its normalized phonetic-distance interval is 0.1--0.5. Acoustic matching uses the available phoneme vocabulary to select distractors.

The following training instructions are selected from the archived pools and presented to the speech model with audio.
\begin{lstlisting}[style=granaryprompt]
Punctuation:
Transcribe the audio with proper punctuation and capitalization.

Spoken form:
Transcribe the audio as plain normalized text, without punctuation or capitalization.

Written form:
Transcribe the audio in written form: render numbers, currencies, and dates using digits and symbols.

Written form without disfluencies:
Transcribe the audio cleanly: remove filler words ('um', 'uh', 'you know'), repetitions, and false starts. Also apply inverse text normalization (digits, symbols).

Caption:
Provide a short caption that describes what this audio is about.

Code switching:
Transcribe the audio, preserving any English words in their original Latin spelling and the rest in the native script.

Translation:
This is {source_lang} speech. Translate it into {target_lang}.

Coarse context:
This audio belongs to the {domain} field. Transcribe the audio into text, ensuring all punctuation marks are included.

Fine context:
This audio belongs to the {domain} field and may contain the following words or phrases: {entities}. Transcribe the audio into text, ensuring all punctuation marks are included.
\end{lstlisting}

The full pools are in \path|prompts/training_instruction_pools.md| and \path|prompts/context_instruction_pools.md|. Each target must match the requested format. For example, an instruction for lowercase text without punctuation requires a lowercase, unpunctuated target.

\clearpage
\section{Reported Downstream Results and Evaluation Instructions}
\label{app:downstream-results}

Table~\ref{tab:transcribe-detailed} presents per-language results from the \transcribename{} technical report~\citep{nemotron35-transcribe}. We use the report's language-specific WER and CER scoring, translation directions, and target tokenizers.

\input{tables/transcribe_detailed}

The reported ASR evaluation disables reasoning and uses the following base instruction:
\begin{lstlisting}[style=granaryprompt]
Audio transcription. Language: as spoken. Wording: verbatim, including fillers and repetitions. Formatting: punctuation and capitalization. Output: transcript only.
\end{lstlisting}
For FLEURS, including its multilingual leaderboard subsets, the report uses:
\begin{lstlisting}[style=granaryprompt]
Audio transcription. Language: as spoken, in its standard writing system. Wording: verbatim, including fillers and repetitions. Formatting: punctuation and capitalization; numbers in their standard written form. Output: transcript only.
\end{lstlisting}
\noindent\begin{minipage}{\linewidth}
The speech translation instruction is:
\begin{lstlisting}[style=granaryprompt]
Please translate the given speech to {target language}.
\end{lstlisting}
\end{minipage}
During Transcribe evaluation, each instruction precedes the audio in the model's chat template.

%% file: tables/transcribe_detailed.tex
\begin{table}[H]
\centering
\scriptsize
\setlength{\tabcolsep}{4pt}
\caption{Transcribe per-language results reproduced from its technical
report~\citep{nemotron35-transcribe}. ASR is WER or CER$^\dagger$ (\%);
lower is better. Translation is BLEU, or spBLEU$^\ast$ for Thai targets;
higher is better. Dashes mark missing benchmark combinations. Norwegian
is retained as reported but is not officially supported for transcription.}
\label{tab:transcribe-detailed}
\begin{tabular}{lrrrrrr}
\toprule
 & \multicolumn{2}{c}{ASR} & \multicolumn{2}{c}{FLEURS translation} & \multicolumn{2}{c}{CoVoST 2 translation} \\
\cmidrule(lr){2-3}\cmidrule(lr){4-5}\cmidrule(lr){6-7}
Language & FLEURS & CoVoST 2 & X$\rightarrow$En & En$\rightarrow$X & X$\rightarrow$En & En$\rightarrow$X \\
\midrule
Arabic & 6.23 & 12.17 & 30.01 & 21.77 & 42.61 & 21.34 \\
Bulgarian & 7.92 & -- & 35.09 & 34.00 & -- & -- \\
Chinese & 8.73$^\dagger$ & 9.67$^\dagger$ & 21.88 & 38.51 & 20.54 & 38.45 \\
Croatian & 8.43 & -- & 32.80 & 25.98 & -- & -- \\
Czech & 5.13 & -- & 35.07 & 27.33 & -- & -- \\
Danish & 9.06 & -- & 38.70 & 39.32 & -- & -- \\
Dutch & 3.81 & 5.37 & 29.74 & 25.47 & 44.34 & -- \\
English & 3.31 & 6.42 & -- & -- & -- & -- \\
Estonian & 9.09 & 15.03 & 31.14 & 21.62 & 20.55 & 24.71 \\
Filipino & 17.93 & -- & 6.40 & 9.46 & -- & -- \\
Finnish & 6.24 & -- & 28.57 & 20.93 & -- & -- \\
French & 3.24 & 4.77 & 38.19 & 42.51 & 41.14 & -- \\
German & 2.71 & 2.89 & 39.96 & 33.93 & 39.44 & 34.29 \\
Greek & 27.09 & -- & 24.44 & 23.78 & -- & -- \\
Hebrew & 20.36 & -- & 28.46 & 25.47 & -- & -- \\
Hindi & 7.50 & -- & 29.62 & 22.09 & -- & -- \\
Hungarian & 7.70 & -- & 28.33 & 20.69 & -- & -- \\
Indonesian & 14.71 & 43.85 & 8.83 & 18.44 & 6.95 & 15.39 \\
Italian & 1.39 & 3.03 & 29.10 & 27.14 & 38.77 & -- \\
Japanese & 4.51$^\dagger$ & 10.05$^\dagger$ & 23.00 & 29.73 & 28.93 & 29.92 \\
Korean & 2.79$^\dagger$ & -- & 24.23 & 27.68 & -- & -- \\
Latvian & 8.19 & 18.99 & 31.06 & 25.60 & 25.69 & 23.60 \\
Lithuanian & 12.66 & -- & 25.44 & 21.03 & -- & -- \\
Maltese & 16.28 & -- & 31.76 & 23.51 & -- & -- \\
Norwegian & 63.26 & -- & 29.74 & 17.24 & -- & -- \\
Polish & 3.76 & -- & 26.54 & 18.56 & -- & -- \\
Portuguese & 3.05 & 2.85 & 44.33 & 42.65 & 54.46 & -- \\
Romanian & 6.02 & -- & 36.76 & 34.01 & -- & -- \\
Russian & 3.72 & 2.34 & 32.73 & 25.83 & 49.31 & -- \\
Slovak & 5.27 & -- & 33.72 & 26.48 & -- & -- \\
Slovenian & 11.13 & 32.97 & 26.83 & 23.54 & 22.46 & 29.49 \\
Spanish & 2.20 & 2.82 & 28.70 & 25.60 & 41.70 & -- \\
Swedish & 7.29 & 14.67 & 39.79 & 38.61 & 40.51 & 38.97 \\
Thai & 7.19$^\dagger$ & -- & 20.19 & 39.20$^\ast$ & -- & -- \\
Turkish & 6.88 & 19.37 & 23.66 & 8.70 & 21.31 & 7.20 \\
Ukrainian & 5.98 & -- & 34.12 & 24.18 & -- & -- \\
Vietnamese & 2.88$^\dagger$ & -- & 22.21 & 24.81 & -- & -- \\
\bottomrule
\end{tabular}
\end{table}

%% file: sections/appendix_text_models.tex
\section{Qualitative Text-Model Comparison}
\label{app:text-model-comparison}

We compared Qwen3.6-27B-FP8, Qwen3.5-35B-A3B-FP8, and Qwen3.5-122B-A10B-FP8~\citep{qwen36-27b,qwen35}. The comparison covered punctuation, capitalization, and inverse text normalization. Each model received the same English AMI, Dutch MOSEL, and Russian YouTube-Commons samples. Table~\ref{tab:text-model-ablation} summarizes their observed editing behavior.

\begin{table}[htbp]
\centering
\caption{Qualitative behavior in the text-model comparison. The column names give model parameter counts in billions.}
\label{tab:text-model-ablation}
\small
\begin{tabularx}{\linewidth}{@{}p{0.25\linewidth}YYY@{}}
\toprule
Behavior & 35B & 27B & 122B \\
\midrule
Dash normalization
  & Uses em dashes
  & Preserves hyphens
  & Uses em dashes \\
Fragment endings
  & Usually preserves fragments
  & Frequently adds periods
  & Mixed behavior \\
Number normalization
  & Conservative
  & Aggressive
  & Moderate \\
Commas after interjections
  & Inconsistent
  & Moderate
  & Most frequent \\
Spelling and ASR correction
  & Actively corrects
  & Usually preserves input
  & Usually preserves input \\
Cyrillic transliteration
  & Usually preserves Cyrillic
  & Occasional
  & Most frequent \\
Quotation marks
  & Adds when inferred
  & Usually preserves input
  & Usually preserves input \\
Sentence joining
  & More frequent
  & Rare
  & Rare \\
\bottomrule
\end{tabularx}
\end{table}

The 35B model changed the text most often. It corrected apparent spelling and recognition errors, joined fragments, added quotation marks, and changed typography. Such edits risk changing the spoken content represented by the transcript.

The 27B model usually retained the input words. It added more punctuation and converted more number expressions into digits. It sometimes added a period to an incomplete fragment. One error converted an expression meaning ``half a day'' to \texttt{1/2}.

The 122B model made fewer spelling and sentence changes than the 35B model. It used more em dashes and commas around interjections than the 27B model. It also converted Russian brand names and abbreviations into Latin script more often.

We selected Qwen3.6-27B-FP8 based on its observed word preservation and processing cost. The comparison also identified number-conversion and fragment-punctuation errors, which motivate task-specific prompts and validation.

%% file: references.bib
@misc{nemo-curator,
  author       = {{NVIDIA}},
  title        = {{NeMo Curator}},
  year         = {2026},
  howpublished = {\url{https://github.com/NVIDIA/NeMo-Curator}},
  note         = {Accessed September 2026}
}

@misc{ray,
  author       = {{The Ray Team}},
  title        = {Ray: A Distributed Computing Framework},
  year         = {2026},
  howpublished = {\url{https://www.ray.io/}},
  note         = {Accessed September 2026}
}

@misc{slurm,
  author       = {{SchedMD}},
  title        = {{Slurm Workload Manager}},
  year         = {2026},
  howpublished = {\url{https://slurm.schedmd.com/}},
  note         = {Accessed September 2026}
}

@misc{vllm,
  author       = {Kwon, Woosuk and Li, Zhuohan and Zhuang, Siyuan and Sheng, Ying and Zheng, Lianmin and Yu, Cody Hao and Gonzalez, Joseph and Zhang, Hao and Stoica, Ion},
  title        = {Efficient Memory Management for Large Language Model Serving with PagedAttention},
  year         = {2023},
  howpublished = {\url{https://arxiv.org/abs/2309.06180}}
}

@misc{dynamo,
  author       = {{NVIDIA}},
  title        = {{NVIDIA Dynamo}},
  year         = {2026},
  howpublished = {\url{https://github.com/ai-dynamo/dynamo}},
  note         = {Accessed September 2026}
}

@inproceedings{granary-v1,
  author    = {Rao Koluguri, Nithin and Sekoyan, Monica and Zelenfroynd, George and
               Meister, Sasha and Ding, Shuoyang and Ginsburg, Boris and others},
  title     = {Granary: Speech Recognition and Translation Dataset in 25 {European} Languages},
  booktitle = {Proc. Interspeech},
  pages     = {3923--3927},
  year      = {2025},
  doi       = {10.21437/Interspeech.2025-540},
  note      = {arXiv:2505.13404}
}

@article{emilia,
  author  = {He, Haorui and Shang, Zengqiang and Wang, Chaoren and Li, Xuyuan and
             Gu, Yicheng and Hua, Hua and Liu, Liwei and Yang, Chen and Li, Jiaqi and
             Shi, Peiyang and Wang, Yuancheng and Chen, Kai and Zhang, Pengyuan and
             Wu, Zhizheng},
  title   = {{Emilia}: A Large-Scale, Extensive, Multilingual, and Diverse Dataset
             for Speech Generation},
  journal = {IEEE/ACM Transactions on Audio, Speech, and Language Processing},
  volume  = {33},
  pages   = {4044--4058},
  year    = {2025},
  note    = {arXiv:2501.15907; extends the SLT 2024 paper arXiv:2407.05361}
}

@inproceedings{gigaspeech2,
  author    = {Yang, Yifan and Song, Zheshu and Zhuo, Jianheng and Cui, Mingyu and
               Li, Jinpeng and Yang, Bo and Du, Yexing and Ma, Ziyang and Liu, Xunying and
               Wang, Ziyuan and Li, Ke and Fan, Shuai and Yu, Kai and Zhang, Wei-Qiang and
               Chen, Guoguo and Chen, Xie},
  title     = {{GigaSpeech~2}: An Evolving, Large-Scale and Multi-domain {ASR} Corpus for
               Low-Resource Languages with Automated Crawling, Transcription and Refinement},
  booktitle = {Proc. Annual Meeting of the Association for Computational Linguistics (ACL)},
  pages     = {2673--2686},
  year      = {2025},
  note      = {arXiv:2406.11546}
}

@inproceedings{yodas,
  author    = {Li, Xinjian and Takamichi, Shinnosuke and Saeki, Takaaki and
               Chen, William and Shiota, Sayaka and Watanabe, Shinji},
  title     = {{YODAS}: {YouTube}-Oriented Dataset for Audio and Speech},
  booktitle = {Proc. IEEE Automatic Speech Recognition and Understanding Workshop (ASRU)},
  year      = {2023},
  doi       = {10.1109/ASRU57964.2023.10389689},
  note      = {arXiv:2406.00899}
}

@inproceedings{yodas-v3,
  author    = {Chen, William and Takamichi, Shinnosuke and Shiota, Sayaka and
               Fukayama, Satoru and Cornell, Samuele and Watanabe, Shinji},
  title     = {{YODAS v3}: Over 1 Million Hours of High-Bandwidth,
               Stereophonic, Multilingual Speech},
  booktitle = {Proc. Interspeech},
  pages     = {5367--5372},
  year      = {2026},
  doi       = {10.21437/Interspeech.2026-386},
  url       = {https://arxiv.org/abs/2609.29448}
}

@inproceedings{mosel,
  author    = {Gaido, Marco and Papi, Sara and Bentivogli, Luisa and Brutti, Alessio and
               Cettolo, Mauro and Gretter, Roberto and Matassoni, Marco and Nabih, Mohamed and
               Negri, Matteo},
  title     = {{MOSEL}: 950,000 Hours of Speech Data for Open-Source Speech Foundation
               Model Training on {EU} Languages},
  booktitle = {Proc. Conference on Empirical Methods in Natural Language Processing (EMNLP)},
  pages     = {13934--13947},
  year      = {2024},
  note      = {arXiv:2410.01036}
}

@inproceedings{voxpopuli,
  author    = {Wang, Changhan and Rivi{\`e}re, Morgane and Lee, Ann and Wu, Anne and
               Talnikar, Chaitanya and Haziza, Daniel and Williamson, Mary and
               Pino, Juan and Dupoux, Emmanuel},
  title     = {{VoxPopuli}: A Large-Scale Multilingual Speech Corpus for Representation
               Learning, Semi-Supervised Learning and Interpretation},
  booktitle = {Proc. Annual Meeting of the Association for Computational Linguistics (ACL)},
  pages     = {993--1003},
  year      = {2021},
  doi       = {10.18653/v1/2021.acl-long.80},
  note      = {arXiv:2101.00390}
}

@inproceedings{mls,
  author    = {Pratap, Vineel and Xu, Qiantong and Sriram, Anuroop and
               Synnaeve, Gabriel and Collobert, Ronan},
  title     = {{MLS}: A Large-Scale Multilingual Dataset for Speech Research},
  booktitle = {Proc. Interspeech},
  pages     = {2757--2761},
  year      = {2020},
  note      = {arXiv:2012.03411}
}

@inproceedings{commonvoice,
  author    = {Ardila, Rosana and Branson, Megan and Davis, Kelly and Kohler, Michael and
               Meyer, Josh and Henretty, Michael and Morais, Reuben and Saunders, Lindsay and
               Tyers, Francis and Weber, Gregor},
  title     = {{Common Voice}: A Massively-Multilingual Speech Corpus},
  booktitle = {Proc. Language Resources and Evaluation Conference (LREC)},
  pages     = {4218--4222},
  year      = {2020},
  note      = {arXiv:1912.06670}
}

@inproceedings{peoples-speech,
  author    = {Galvez, Daniel and Diamos, Greg and Torres, Juan and Achorn, Keith and
               Cer{\'o}n, Juan and Gopi, Anjali and Kanter, David and Lam, Max and
               Mazumder, Mark and Reddi, Vijay Janapa},
  title     = {The {People's Speech}: A Large-Scale Diverse {English} Speech Recognition
               Dataset for Commercial Usage},
  booktitle = {Proc. NeurIPS Track on Datasets and Benchmarks},
  year      = {2021},
  note      = {arXiv:2111.09344}
}

@inproceedings{librilight,
  author    = {Kahn, Jacob and Rivi{\`e}re, Morgane and Zheng, Weiyi and
               Kharitonov, Evgeny and Xu, Qiantong and Mazar{\'e}, Pierre-Emmanuel and
               Karadayi, Julien and Liptchinsky, Vitaliy and Collobert, Ronan and
               Fuegen, Christian and Likhomanenko, Tatiana and Synnaeve, Gabriel and
               Joulin, Armand and Mohamed, Abdelrahman and Dupoux, Emmanuel},
  title     = {{Libri-Light}: A Benchmark for {ASR} with Limited or No Supervision},
  booktitle = {Proc. IEEE International Conference on Acoustics, Speech and Signal
               Processing (ICASSP)},
  pages     = {7669--7673},
  year      = {2020},
  note      = {arXiv:1912.07875}
}

@misc{youtube-commons,
  author       = {{PleIAs}},
  title        = {{YouTube-Commons}: A Corpus of {Creative Commons} {YouTube} Audio and Transcripts},
  year         = {2024},
  howpublished = {\url{https://huggingface.co/datasets/PleIAs/YouTube-Commons}},
  note         = {Accessed September 2026}
}

@misc{commonvoice-v26,
  author       = {{MCV}},
  title        = {{Common Voice Scripted Speech 26.0}},
  year         = {2026},
  howpublished = {\url{https://github.com/common-voice/cv-dataset/blob/main/datasets/scripted-speech/cv-corpus-26.0-2026-06-12.json}},
  note         = {Released June 2026; accessed September 2026}
}

@misc{autoprep,
  author       = {Yu, Jianwei and Chen, Hangting and Bian, Yanyao and Li, Xiang and
                  Luo, Yi and Tian, Jinchuan and Liu, Mengyang and Jiang, Jiayi and
                  Wang, Shuai},
  title        = {{AutoPrep}: An Automatic Preprocessing Framework for In-the-Wild
                  Speech Data},
  year         = {2023},
  howpublished = {arXiv preprint arXiv:2309.13905}
}

@misc{transcription-policy,
  author       = {Wagner, Laurin and Zusag, Mario and Thallinger, Bernhard},
  title        = {Transcription Policy as a Latent Variable: Activating
                  Controllable Verbatim {ASR} with Word-Level Timing},
  year         = {2026},
  howpublished = {arXiv preprint arXiv:2607.18934}
}

@inproceedings{conec,
  author    = {Huang, Ruizhe and Yarmohammadi, Mahsa and Trmal, Jan and
               Liu, Jing and Raj, Desh and Garcia, Leibny Paola and
               Ivanov, Alexei and Ehlen, Patrick and Yu, Mingzhi and
               Rastrow, Ariya and Povey, Dan and Khudanpur, Sanjeev},
  title     = {{ConEC}: Earnings Call Dataset with Real-World Contexts
               for Benchmarking Contextual Speech Recognition},
  booktitle = {Proceedings of the 2024 Joint International Conference
               on Computational Linguistics, Language Resources and
               Evaluation},
  pages     = {3700--3706},
  year      = {2024},
  url       = {https://aclanthology.org/2024.lrec-main.328/}
}

@misc{covost2,
  author       = {Wang, Changhan and Wu, Anne and Pino, Juan},
  title        = {{CoVoST~2} and Massively Multilingual
                  Speech-to-Text Translation},
  year         = {2020},
  howpublished = {arXiv preprint arXiv:2007.10310}
}

@misc{speechverse,
  author       = {Das, Nilaksh and Dingliwal, Saket and Ronanki, Srikanth and
                  Paturi, Rohit and Huang, Zhaocheng and Mathur, Prashant and
                  Yuan, Jie and Bekal, Dhanush and Niu, Xing and
                  Jayanthi, Sai Muralidhar and Li, Xilai and Mundnich, Karel and
                  Sunkara, Monica and Srinivasan, Sundararajan and
                  Garcia-Romero, Daniel and Han, Kyu J. and Kirchhoff, Katrin},
  title        = {{SpeechVerse}: A Large-Scale Generalizable
                  Audio-Language Model},
  year         = {2024},
  howpublished = {arXiv preprint arXiv:2405.08295}
}

@misc{streaming-sortformer,
  author       = {Medennikov, Ivan and Park, Taejin and Wang, Weiqing and
                  Huang, He and Dhawan, Kunal and Wang, Jinhan and
                  Balam, Jagadeesh and Ginsburg, Boris},
  title        = {Streaming {Sortformer}: Speaker Cache-Based Online
                  Speaker Diarization with Arrival-Time Ordering},
  year         = {2025},
  howpublished = {arXiv preprint arXiv:2507.18446},
  note         = {Accepted to Interspeech 2025}
}

@misc{nemotron3-diarization,
  author       = {{NVIDIA}},
  title        = {{Nemotron 3 Diarization}},
  year         = {2026},
  howpublished = {\url{https://huggingface.co/nvidia/Nemotron-3-Diarization}},
  note         = {Hugging Face model card; accessed 20 October 2026}
}

@article{panns,
  author  = {Kong, Qiuqiang and Cao, Yin and Iqbal, Turab and Wang, Yuxuan and Wang, Wenwu and Plumbley, Mark D.},
  title   = {{PANNs}: Large-Scale Pretrained Audio Neural Networks for Audio Pattern Recognition},
  journal = {IEEE/ACM Transactions on Audio, Speech, and Language Processing},
  volume  = {28},
  pages   = {2880--2894},
  year    = {2020},
  doi     = {10.1109/TASLP.2020.3030497},
  note    = {arXiv:1912.10211}
}

@misc{silero-vad,
  author       = {{Silero Team}},
  title        = {{Silero VAD}: Pre-Trained Enterprise-Grade Voice Activity Detector, Number Detector, and Language Classifier},
  year         = {2024},
  howpublished = {\url{https://github.com/snakers4/silero-vad}},
  note         = {Software repository; accessed September 2026}
}

@inproceedings{voxlingua107,
  author    = {Valk, J{\"o}rgen and Alum{\"a}e, Tanel},
  title     = {{VoxLingua107}: A Dataset for Spoken Language Recognition},
  booktitle = {Proceedings of the IEEE Spoken Language Technology Workshop},
  year      = {2021}
}

@misc{speechbrain-voxlingua107-ecapa,
  author       = {{SpeechBrain}},
  title        = {{VoxLingua107 ECAPA-TDNN} Spoken Language Identification Model},
  year         = {n.d.},
  howpublished = {\url{https://huggingface.co/speechbrain/lang-id-voxlingua107-ecapa}},
  note         = {Hugging Face model card; accessed September 2026}
}

@misc{qwen36-27b,
  author = {{Qwen Team}},
  title  = {{Qwen3.6-27B}: Flagship-Level Coding in a {27B} Dense Model},
  month  = {April},
  year   = {2026},
  url    = {https://qwen.ai/blog?id=qwen3.6-27b}
}

@misc{qwen35,
  author = {{Qwen Team}},
  title  = {{Qwen3.5}: Towards Native Multimodal Agents},
  month  = {February},
  year   = {2026},
  url    = {https://qwen.ai/blog?id=qwen3.5}
}

@inproceedings{srivastav2025openasr,
  title     = {Open {ASR} Leaderboard: Towards Reproducible and Transparent Multilingual and Long-Form Speech Recognition Evaluation},
  author    = {Srivastav, Vaibhav and Zheng, Steven and Bezzam, Eric and Le Bihan, Eustache and Koluguri, Nithin Rao and {\.Z}elasko, Piotr and Majumdar, Somshubra and Moumen, Adel and Gandhi, Sanchit},
  booktitle = {Proc. Interspeech},
  pages     = {6692--6697},
  year      = {2026},
  doi       = {10.21437/Interspeech.2026-1902}
}

@inproceedings{chen2023salm,
  title     = {{SALM}: Speech-augmented Language Model with In-context Learning for Speech Recognition and Translation},
  author    = {Chen, Zhehuai and Huang, He and Andrusenko, Andrei and Hrinchuk, Oleksii and Puvvada, Krishna C. and Li, Jason and Ghosh, Subhankar and Balam, Jagadeesh and Ginsburg, Boris},
  booktitle = {Proc. IEEE International Conference on Acoustics, Speech and Signal Processing (ICASSP)},
  pages     = {13521--13525},
  year      = {2024},
  doi       = {10.1109/ICASSP48485.2024.10447553}
}

@techreport{nemotron35-transcribe,
  author = {{NVIDIA}},
  title = {{Nemotron 3.5 Transcribe}: Technical Report},
  institution = {NVIDIA},
  year = {2026},
  url = {https://huggingface.co/nvidia/NVIDIA-Nemotron-3.5-Transcribe-30B-A4B}
}

@misc{granary-v2-data,
  author       = {{NVIDIA}},
  title        = {{Granary-v2}},
  year         = {2026},
  howpublished = {\url{https://huggingface.co/datasets/nvidia/Granary-v2}},
  note         = {Hugging Face dataset repository}
}

@misc{indic-transcribe-core,
  author       = {{Bodhan AI}},
  title        = {{Indic-Transcribe-core}},
  year         = {2026},
  howpublished = {\url{https://huggingface.co/bodhan-ai/indic-transcribe-core}},
  note         = {Hugging Face model card; accessed 21 October 2026}
}

@misc{indicconformer-monolingual,
  author       = {{AI4Bharat}},
  title        = {{IndicConformer Monolingual ASR Models}},
  year         = {n.d.},
  howpublished = {\url{https://huggingface.co/collections/ai4bharat/indicconformer}},
  note         = {Per-language hybrid CTC-RNNT checkpoints; accessed 23 October 2026}
}

@inproceedings{gowda-etal-2023-cometoid,
    title = "Cometoid: Distilling Strong Reference-based Machine Translation Metrics into {E}ven Stronger Quality Estimation Metrics",
    author = "Gowda, Thamme  and
      Kocmi, Tom  and
      Junczys-Dowmunt, Marcin",
    editor = "Koehn, Philipp  and
      Haddon, Barry  and
      Kocmi, Tom  and
      Monz, Christof",
    booktitle = "Proceedings of the Eighth Conference on Machine Translation",
    month = dec,
    year = "2023",
    address = "Singapore",
    publisher = "Association for Computational Linguistics",
    url = "https://aclanthology.org/2023.wmt-1.62",
    pages = "751--755",
}

@article{conneau2022fleurs,
  author = {Conneau, Alexis and Ma, Min and Khanuja, Simran and Zhang, Yu and Axelrod, Vera and Dalmia, Siddharth and Riesa, Jason and Rivera, Clara and Bapna, Ankur},
  title = {{FLEURS}: Few-shot Learning Evaluation of Universal Representations of Speech},
  journal = {arXiv preprint arXiv:2205.12446},
  year = {2022},
  url = {https://arxiv.org/abs/2205.12446}
}

@inproceedings{radford2022whisper,
  title     = {Robust Speech Recognition via Large-Scale Weak Supervision},
  author    = {Radford, Alec and Kim, Jong Wook and Xu, Tao and Brockman, Greg and McLeavey, Christine and Sutskever, Ilya},
  booktitle = {Proc. 40th International Conference on Machine Learning (ICML)},
  series    = {Proceedings of Machine Learning Research},
  volume    = {202},
  pages     = {28492--28518},
  publisher = {PMLR},
  year      = {2023}
}

@misc{systran_fasterwhisper,
  title        = {Faster Whisper Transcription with {CTranslate2}},
  author       = {{SYSTRAN}},
  year         = {n.d.},
  howpublished = {\url{https://github.com/SYSTRAN/faster-whisper}},
  note         = {Software repository}
}

@misc{voicearena2026monsoon,
  author = {{Voice Arena}},
  title = {{MonsoonASR}: Open {ASR} Leaderboard {English} Test Set ({en-IN})},
  year = {2026},
  howpublished = {\url{https://huggingface.co/datasets/VoiceArena/MonsoonASR-Open-ASR-leaderboard-en-IN}},
  note = {Hugging Face dataset card; accessed 26 October 2026}
}

@article{delrio2022earnings22,
  title   = {Earnings-22: A Practical Benchmark for Accents in the Wild},
  author  = {Del Rio, Miguel and Ha, Peter and McNamara, Quinten and Miller, Corey and Chandra, Shipra},
  journal = {arXiv preprint arXiv:2203.15591},
  year    = {2022}
}

@inproceedings{chen2021gigaspeech,
  title     = {{GigaSpeech}: An Evolving, Multi-domain {ASR} Corpus with 10,000 Hours of Transcribed Audio},
  author    = {Chen, Guoguo and Chai, Shuzhou and Wang, Guan-Bo and Du, Jiayu and Zhang, Wei-Qiang and Weng, Chao and Su, Dan and Povey, Daniel and Trmal, Jan and Zhang, Junbo and Jin, Mingjie and Khudanpur, Sanjeev and Watanabe, Shinji and Zhao, Shuaijiang and Zou, Wei and Li, Xiangang and Yao, Xuchen and Wang, Yongqing and You, Zhao and Yan, Zhiyong},
  booktitle = {Proc. Interspeech},
  pages     = {3670--3674},
  year      = {2021},
  doi       = {10.21437/Interspeech.2021-1965}
}

@inproceedings{oneill2021spgispeech,
  title     = {{SPGISpeech}: 5,000 hours of transcribed financial audio for fully formatted end-to-end speech recognition},
  author    = {O'Neill, Patrick K. and Lavrukhin, Vitaly and Majumdar, Somshubra and Noroozi, Vahid and Zhang, Yuekai and Kuchaiev, Oleksii and Balam, Jagadeesh and Dovzhenko, Yuliya and Freyberg, Keenan and Shulman, Michael D. and Ginsburg, Boris and Watanabe, Shinji and Kucsko, Georg},
  booktitle = {Proc. Interspeech},
  pages     = {1434--1438},
  year      = {2021},
  doi       = {10.21437/Interspeech.2021-1860}
}

@inproceedings{zhang2024urgent,
  title     = {{URGENT} Challenge: Universality, Robustness, and Generalizability For Speech Enhancement},
  author    = {Zhang, Wangyou and Scheibler, Robin and Saijo, Kohei and Cornell, Samuele and Li, Chenda and Ni, Zhaoheng and Pirklbauer, Jan and Sach, Marvin and Watanabe, Shinji and Fingscheidt, Tim and Qian, Yanmin},
  booktitle = {Proc. Interspeech},
  pages     = {4868--4872},
  year      = {2024},
  doi       = {10.21437/Interspeech.2024-1239}
}

@inproceedings{carletta2005ami,
  title     = {The {AMI} Meeting Corpus: A Pre-announcement},
  author    = {Carletta, Jean and Ashby, Simone and Bourban, Sebastien and Flynn, Mike and Guillemot, Mael and Hain, Thomas and Kadlec, Jaroslav and Karaiskos, Vasilis and Kraaij, Wessel and Kronenthal, Melissa and Lathoud, Guillaume and Lincoln, Mike and Lisowska, Agnes and McCowan, Iain and Post, Wilfried and Reidsma, Dennis and Wellner, Pierre},
  booktitle = {Machine Learning for Multimodal Interaction (MLMI 2005)},
  series    = {Lecture Notes in Computer Science},
  volume    = {3869},
  pages     = {28--39},
  publisher = {Springer},
  year      = {2006},
  doi       = {10.1007/11677482_3}
}

@inproceedings{panayotov2015librispeech,
  title     = {Librispeech: An {ASR} corpus based on public domain audio books},
  author    = {Panayotov, Vassil and Chen, Guoguo and Povey, Daniel and Khudanpur, Sanjeev},
  booktitle = {Proc. IEEE International Conference on Acoustics, Speech and Signal Processing (ICASSP)},
  pages     = {5206--5210},
  year      = {2015},
  doi       = {10.1109/ICASSP.2015.7178964}
}

@inproceedings{koluguri2026preferenceasr,
  title     = {Preference-{ASR}: A Preference-Aware Test Set for Benchmarking {ASR} in the Era of Speech {LLMs}},
  author    = {Koluguri, Nithin Rao and Meister, Sasha and Karpov, Nikolay and {\.Z}elasko, Piotr and Raj, Desh and Balam, Jagadeesh and Ginsburg, Boris},
  booktitle = {Proc. Interspeech},
  pages     = {6687--6691},
  year      = {2026},
  doi       = {10.21437/Interspeech.2026-728}
}

@article{xu2025qwen3omni,
  title   = {{Qwen3-Omni} Technical Report},
  author  = {Xu, Jin and Guo, Zhifang and Hu, Hangrui and Chu, Yunfei and Wang, Xiong and He, Jinzheng and Wang, Yuxuan and Shi, Xian and He, Ting and Zhu, Xinfa and Lv, Yuanjun and Wang, Yongqi and Guo, Dake and Wang, He and Ma, Linhan and Zhang, Pei and Zhang, Xinyu and Hao, Hongkun and Guo, Zishan and Yang, Baosong and Zhang, Bin and Ma, Ziyang and Wei, Xipin and Bai, Shuai and Chen, Keqin and Liu, Xuejing and Wang, Peng and Yang, Mingkun and Liu, Dayiheng and Ren, Xingzhang and Zheng, Bo and Men, Rui and Zhou, Fan and Yu, Bowen and Yang, Jianxin and Yu, Le and Zhou, Jingren and Lin, Junyang},
  journal = {arXiv preprint arXiv:2509.17765},
  year    = {2025}
}

@article{shi2026qwen3asr,
  title   = {{Qwen3-ASR} Technical Report},
  author  = {Shi, Xian and Wang, Xiong and Guo, Zhifang and Wang, Yongqi and Zhang, Pei and Zhang, Xinyu and Guo, Zishan and Hao, Hongkun and Xi, Yu and Yang, Baosong and Xu, Jin and Zhou, Jingren and Lin, Junyang},
  journal = {arXiv preprint arXiv:2601.21337},
  year    = {2026}
}

@misc{parakeet-v3,
  author = {{NVIDIA}},
  title = {{Parakeet-TDT-0.6B-v3}},
  year = {2025},
  howpublished = {\url{https://huggingface.co/nvidia/parakeet-tdt-0.6b-v3}},
  note = {Hugging Face model card; accessed 26 October 2026}
}

@misc{whisper-large-v3,
  author = {{OpenAI}},
  title = {{Whisper Large v3}},
  year = {2023},
  howpublished = {\url{https://huggingface.co/openai/whisper-large-v3}},
  note = {Hugging Face model card; accessed 26 October 2026}
}

@misc{hebrew-whisper,
  author = {{ivrit.ai}},
  title = {{Hebrew Whisper Large v3}},
  year = {2025},
  howpublished = {\url{https://huggingface.co/ivrit-ai/whisper-large-v3}},
  note = {Hugging Face model card; accessed 26 October 2026}
}

@misc{gemma4-31b,
  author = {{Google}},
  title = {{Gemma 4 31B IT}},
  year = {2026},
  howpublished = {\url{https://huggingface.co/google/gemma-4-31B-it}},
  note = {Hugging Face model card; accessed 26 October 2026}
}

@misc{gemma4-fp8,
  author = {{Red Hat AI}},
  title = {{Gemma-4-31B-it-FP8-dynamic}},
  year = {2026},
  howpublished = {\url{https://huggingface.co/RedHatAI/gemma-4-31B-it-FP8-dynamic}},
  note = {Hugging Face model card; accessed 26 October 2026}
}

@misc{reazonspeech-v2,
  author = {{Reazon Human Interaction Lab}},
  title = {{ReazonSpeech v2}: Japanese Speech Corpus},
  year = {2024},
  howpublished = {\url{https://huggingface.co/datasets/reazon-research/reazonspeech}},
  note = {Hugging Face dataset card; accessed 26 October 2026}
}

@inproceedings{falar,
  author = {Teixeira, Francisco and Carvalho, Carlos and Juli{\~a}o, Mariana and Botelho, Catarina and Solera-Ure{\~n}a, Rub{\'e}n and Paulo, S{\'e}rgio and Rolland, Thomas and Peters, Ben and Trancoso, Isabel and Abad, Alberto},
  title = {{FalAR}: A Large-scale Speaker-Annotated {European Portuguese} Speech Corpus of Parliamentary Sessions},
  booktitle = {Proc. International Conference on Language Resources and Evaluation (LREC)},
  year = {2026},
  url = {https://huggingface.co/datasets/inesc-id/FalAR}
}

@inproceedings{national-speech-corpus,
  author = {Koh, Jia Xin and Mislan, Aqilah and Khoo, Kevin and Ang, Brian and Ang, Wilson and Ng, Charmaine and Tan, Ying-Ying},
  title = {Building the {Singapore English National Speech Corpus}},
  booktitle = {Proc. Interspeech},
  year = {2019},
  pages = {321--325},
  doi = {10.21437/Interspeech.2019-1525}
}

@article{golos,
  author = {Karpov, Nikolay and Denisenko, Alexander and Minkin, Fedor},
  title = {{Golos}: {Russian} Dataset for Speech Research},
  journal = {arXiv preprint arXiv:2106.10161},
  year = {2021},
  url = {https://arxiv.org/abs/2106.10161}
}

@inproceedings{europarl-asr,
  author = {Garc{\'e}s D{\'i}az-Mun{\'i}o, Gon{\c c}al V. and Silvestre-Cerd{\`a}, Joan Albert and Jorge, Javier and Gim{\'e}nez, Adri{\`a} and Iranzo-S{\'a}nchez, Javier and Baquero-Arnal, Pau and Rosell{\'o}, Nahuel and P{\'e}rez-Gonz{\'a}lez-de-Martos, Alejandro and Civera, Jorge and Sanchis, Albert and Juan, Alfons},
  title = {{Europarl-ASR}: A Large Corpus of Parliamentary Debates for Streaming {ASR} Benchmarking and Speech Data Filtering/Verbatimization},
  booktitle = {Proc. Interspeech},
  year = {2021},
  pages = {3695--3699},
  doi = {10.21437/Interspeech.2021-1905}
}

@article{librispeech-pc,
  author = {Meister, Aleksandr and Novikov, Matvei and Karpov, Nikolay and Bakhturina, Evelina and Lavrukhin, Vitaly and Ginsburg, Boris},
  title = {{LibriSpeech-PC}: Benchmark for Evaluation of Punctuation and Capitalization Capabilities of End-to-End {ASR} Models},
  journal = {arXiv preprint arXiv:2310.02943},
  year = {2023},
  url = {https://arxiv.org/abs/2310.02943}
}

@article{dusha,
  author = {Kondratenko, Vladimir and Sokolov, Artem and Karpov, Nikolay and Kutuzov, Oleg and Savushkin, Nikita and Minkin, Fyodor},
  title = {Large Raw Emotional Dataset with Aggregation Mechanism},
  journal = {arXiv preprint arXiv:2212.12266},
  year = {2022},
  url = {https://arxiv.org/abs/2212.12266}
}

@misc{vctk,
  author = {Yamagishi, Junichi and Veaux, Christophe and MacDonald, Kirsten},
  title = {{CSTR VCTK Corpus}: {English} Multi-speaker Corpus for {CSTR} Voice Cloning Toolkit (Version 0.92)},
  year = {2019},
  howpublished = {University of Edinburgh, Centre for Speech Technology Research},
  doi = {10.7488/ds/2645},
  url = {https://datashare.ed.ac.uk/handle/10283/3443}
}

@article{speech-commands,
  author = {Warden, Pete},
  title = {Speech Commands: A Dataset for Limited-Vocabulary Speech Recognition},
  journal = {arXiv preprint arXiv:1804.03209},
  year = {2018},
  url = {https://arxiv.org/abs/1804.03209}
}

@inproceedings{tdt,
  author = {Xu, Hainan and Jia, Fei and Majumdar, Somshubra and Huang, He and Watanabe, Shinji and Ginsburg, Boris},
  title = {Efficient Sequence Transduction by Jointly Predicting Tokens and Durations},
  booktitle = {Proc. 40th International Conference on Machine Learning},
  series = {Proceedings of Machine Learning Research},
  volume = {202},
  year = {2023},
  pages = {38462--38484},
  url = {https://proceedings.mlr.press/v202/xu23g.html}
}

@inproceedings{comet,
  author = {Rei, Ricardo and Stewart, Craig and Farinha, Ana C and Lavie, Alon},
  title = {{COMET}: A Neural Framework for {MT} Evaluation},
  booktitle = {Proc. Conference on Empirical Methods in Natural Language Processing (EMNLP)},
  year = {2020},
  pages = {2685--2702},
  doi = {10.18653/v1/2020.emnlp-main.213},
  url = {https://aclanthology.org/2020.emnlp-main.213/}
}

@inproceedings{bleu,
  author = {Papineni, Kishore and Roukos, Salim and Ward, Todd and Zhu, Wei-Jing},
  title = {{BLEU}: A Method for Automatic Evaluation of Machine Translation},
  booktitle = {Proc. 40th Annual Meeting of the Association for Computational Linguistics},
  year = {2002},
  pages = {311--318},
  doi = {10.3115/1073083.1073135},
  url = {https://aclanthology.org/P02-1040/}
}

@inproceedings{sacrebleu,
  author = {Post, Matt},
  title = {A Call for Clarity in Reporting {BLEU} Scores},
  booktitle = {Proc. Third Conference on Machine Translation: Research Papers},
  year = {2018},
  pages = {186--191},
  doi = {10.18653/v1/W18-6319},
  url = {https://aclanthology.org/W18-6319/}
}

@article{flores101,
  author = {Goyal, Naman and Gao, Cynthia and Chaudhary, Vishrav and Chen, Peng-Jen and Wenzek, Guillaume and Ju, Da and Krishnan, Sanjana and Ranzato, Marc'Aurelio and Guzm{\'a}n, Francisco and Fan, Angela},
  title = {The {FLORES-101} Evaluation Benchmark for Low-Resource and Multilingual Machine Translation},
  journal = {Transactions of the Association for Computational Linguistics},
  volume = {10},
  year = {2022},
  pages = {522--538},
  doi = {10.1162/tacl_a_00474},
  url = {https://aclanthology.org/2022.tacl-1.30/}
}

@article{nllb,
  author = {{NLLB Team} and Costa-juss{\`a}, Marta R. and Cross, James and {\c C}elebi, Onur and Elbayad, Maha and Heafield, Kenneth and Heffernan, Kevin and others},
  title = {No Language Left Behind: Scaling Human-Centered Machine Translation},
  journal = {arXiv preprint arXiv:2207.04672},
  year = {2022},
  url = {https://arxiv.org/abs/2207.04672}
}

@article{sekoyan2025canaryv2,
  title   = {{Canary-1B-v2} \& {Parakeet-TDT-0.6B-v3}: Efficient and High-Performance Models for Multilingual {ASR} and {AST}},
  author  = {Sekoyan, Monica and Koluguri, Nithin Rao and Tadevosyan, Nune and {\.Z}elasko, Piotr and Bartley, Travis and Karpov, Nikolay and Balam, Jagadeesh and Ginsburg, Boris},
  journal = {arXiv preprint arXiv:2509.14128},
  year    = {2025},
  url     = {https://arxiv.org/abs/2509.14128}
}
